\documentclass[fleqn,usenatbib]{mnras}

\usepackage[T1]{fontenc}

\DeclareRobustCommand{\VAN}[3]{#2}
\let\VANthebibliography\thebibliography
\def\thebibliography{\DeclareRobustCommand{\VAN}[3]{##3}\VANthebibliography}

\usepackage{graphicx}	
\usepackage{amsmath}	
\usepackage{amssymb}	

\usepackage[dvipsnames]{xcolor}
\usepackage{subcaption}

\usepackage{multirow}

\newcommand{\usmg}{USMg\textsc{ii}}
  
\newcommand{\lya}{Ly$\alpha$ }

\newcommand{\zabs}{$z_{abs}$ }
\newcommand{\kms}{$km s^{-1}$ }

\newcommand{\HI}{\mbox{H\,{\sc i}}}

\newcommand{\OII}{[\mbox{O\,{\sc ii}}]}
\newcommand{\OIII}{[\mbox{O\,{\sc iii}}]}

\newcommand{\MgII}{\mbox{Mg\,{\sc ii}}}
\newcommand{\MgI}{\mbox{Mg\,{\sc i}}}

\newcommand{\FeII}{\mbox{Fe\,{\sc ii}}}
\newcommand{\MnII}{\mbox{Mn\,{\sc ii}}}
\newcommand{\TiII}{\mbox{Ti\,{\sc ii}}}
\newcommand{\CaII}{\mbox{Ca\,{\sc ii}}}

\newcommand{\NeIII}{\mbox{[Ne\,{\sc iii}]}}

\title[Nature of GOTOQ]{Nature of the Galaxies On Top Of Quasars producing \MgII\ absorption}

\author[Guha et al.]{
Labanya Kumar Guha,$^{1}$\thanks{E-mail: labanya@iucaa.in (LG)}
Raghunathan Srianand $^{1}$ \thanks{E-mail: anand@iucaa.in (RS)}
\\
$^{1}$Inter-University Centre for Astronomy \& Astrophysics, Pune, 411007, India
}

\date{Accepted XXX. Received YYY; in original form ZZZ}

\pubyear{2022}

\begin{document}
\label{firstpage}
\pagerange{\pageref{firstpage}--\pageref{lastpage}}
\maketitle

\begin{abstract}
Quasar-galaxy pairs at small separations are important probes of gas flows in the disk-halo interface in galaxies. We study host galaxies of 198 \MgII\ absorbers at $0.39\le z_{abs}\le1.05$ that show detectable nebular emission lines in the SDSS spectra. We report measurements of  impact parameter (5.9$\le D[kpc]\le$16.9) and absolute B-band magnitude ($-18.7\le {\rm M_B}\le -22.3$ mag) of host galaxies of 74 of these absorbers using multi-band images from the DESI Legacy Imaging Survey, more than doubling
the number of known host galaxies with $D\le17$ kpc. This has allowed us to quantify the relationship between \MgII\ rest equivalent width($W_{2796}$) and D, with best fit parameters of $W_{2796}(D=0) = 3.44\pm 0.20$\AA\ and an exponential scale length of 21.6$^{+2.41}_{-1.97}$ $kpc$ . We find a significant anti-correlation between $M_B$ and D, and $M_B$ and  $W_{2796}$, consistent with the brighter galaxies producing stronger \MgII\ absorption. We use stacked images to detect average emissions from galaxies in the full sample. Using these images and stacked spectra, we derive the mean stellar mass ($9.4\le log(M_*/M_\odot) \le 9.8$), star formation rate ($2.3\le{\rm SFR}[M_\odot yr^{-1}] \le 4.5$), age (2.5$-$4 Gyr), metallicity (12+log(O/H)$\sim$8.3) and ionization parameter (log~q[cm s$^{-1}$]$\sim$ 7.7) for these galaxies. The average $M_*$ found is less than that of \MgII\ absorbers studied in the literature.  The average SFR and metallicity inferred are consistent with that expected in the main sequence and the known stellar mass-metallicity relation, respectively.  High spatial resolution follow-up spectroscopic and imaging observations of this sample are imperative for probing gas flows close to the star-forming regions of high-$z$ galaxies.
\end{abstract}




\section{Introduction}
\label{sec:introduction}

Galaxies evolve through a slowly varying equilibrium between the gas inflows from the intergalactic medium (IGM), gas outflows from the galaxy to the circumgalactic medium (CGM) and the IGM, and insitu star-formation occurring within the galaxy \citep{Erb2008, Kennicutt2012, Putman2012, Kacprzak2017_gas_accretion, tumlinson2013}. This is usually referred to as the ``baryonic cycle''. Obtaining direct constraints on the gas inflow and outflow rates and how they evolve with redshift (i.e., cosmic time) is very important for our understanding of galaxy evolution.

Large-scale galactic outflows throw materials out to large distances and usually provide negative feedback to the ongoing star formation by preventing further gas accretion and/or not allowing the gas to cool down \citep{Tumlinson2011}. With time, however, the ejected material may cool and eventually fall back to the galactic disc to sustain subsequent star formation by giving rise to the so-called ``galactic fountain" \citep{shapiro1976consequences, houck1990low}. In some cases, the wind may escape the galaxy and thereby enrich the IGM or the intra-group medium \citep[for example,][]{Samui2008}. The signatures of galactic winds or the recycling of it are, more easily seen at the disk-halo interface, i.e., typically around a few kpc from the galactic disk. 

The ``down-the-barrel''  spectroscopy of high-$z$ galaxies, in principle, allows us to probe the connection between the properties of galaxies and the outflowing gas. Such studies confirm the ubiquitous presence of galactic scale outflows (with velocities of 100-1000 $km\,s^{-1}$), traced by the blue-shifted absorption lines of the neutral or singly ionized species like Na $\textsc{i}$, \MgII, \CaII\ and \FeII, in high redshift galaxies \citep[e.g.][]{Tremonti2007, Martin_2012, Rubin2014, Bordoloi2014}. However, it is difficult to constrain the exact locations of these outflows along our line of sight with respect to the galaxy and accurately measure the quantities like mass outflow rate. Spatially resolved galaxy spectra, either with the help of gravitationally lensed background sources \citep[as in][]{Lopez2018,Mortensen2021} or integral field spectroscopy aided with adaptive optics, will allow us to probe the spatial distribution of gas and its kinematics in more details.

Quasar absorption line studies, on the other hand, provide the physical conditions and the kinematics of the absorbing gas in the disk and/or the CGM of the foreground galaxy along a pencil beam. The \MgII\ $\rm{\lambda\lambda\, 2796, 2803}$ absorption doublet present in the spectra of the background quasars is an excellent probe of cold ($\rm{T \sim 10^{4}\,K }$) low-ionization gas associated with a wide range of neutral hydrogen column densities \citep{Srianand1996,Rigby2002, bond2001high}. 
Thanks to its rest wavelength, \MgII\ absorption from low-$z$ galaxies (i.e., $0.3 \leqslant z \leqslant 1.0$) are easily accessible through ground-based optical spectroscopic observations. It is also well documented that the detection probability of DLAs and \HI\ 21-cm absorption increases with increasing rest equivalent width of \MgII\ absorption \citep[see for example,][]{rao2011,gupta2012,Dutta2017}.

A quasar line of sight passing very close to the center of an intervening galaxy (within an impact parameter, $\rm{D\, \lesssim 15\, kpc}$) is bound to probe the disk-halo interface of the galaxy. Although efforts were made to study the \MgII\  absorption at small impact parameters (i.e., D $\lesssim$ 10 kpc) for some galaxies \citep{Kacprzak_2013}, they are all at low-redshift (i.e., $z<$0.1). One way to find the quasar-galaxy pairs at such low impact parameters at high-$z$ is to search for the host galaxies of Ultra-Strong \MgII\ absorption lines (\usmg, having rest equivalent width of \MgII\ $\lambda\, 2796$ line, $\rm{W_{2796}}\geqslant$ 3\AA) as they are expected to have extremely low impact parameters (D) based on the well known $\rm{W_{2796} - D}$ anti-correlation \citep{Chen_2010, Nielsen2013}. However, it has been found in such absorbers that large equivalent widths not only originate from galaxies at low impact parameters but also from groups of galaxies \citep[][]{Nestor_2011, Gauthier2013, Guha2022}.

Another efficient way of identifying quasar-galaxy pairs with low impact parameters is to search for nebular emission lines from the foreground galaxies in the SDSS fiber spectra of background quasars \citep[Galaxies On Top Of Quasar $-$ GOTOQ, as defined by][]{york2012}. In the case of GOTOQs, a star-forming foreground galaxy is present within an angular separation of $\sim1^{\prime \prime}$ (SDSS DR-12) or $\sim1.5^{\prime\prime}$ (SDSS DR-7) 
and the nebular line emissions from these foreground galaxies are present in the spectra of the background quasar.

Without any prior knowledge of the line of sight absorption, just by searching for nebular emission lines (like H$\alpha$, \OIII\, and \OII) from foreground galaxies in the SDSS quasar spectra, a total of 103 GOTOQ were identified in the redshift range $0 \le z_{abs} \le 0.84 $ \citep{Noterdaeme_2010, york2012,straka2013, Straka_2015}.
Using SDSS photometry, they detect galaxies in about 68\% of the galaxies with impact parameters in the range of 0.37 to 12.68 kpc. As most of them are at $z<0.3$, no information is available on the nature of \MgII\ absorption along the quasar sightlines. For GOTOQs at $z>0.3$, \citet{Noterdaeme_2010} have shown that a  strong \MgII\ absorption is always detected from the GOTOQs. Similarly, based on a very small sample of GOTOQs at $z<0.1$ towards UV bright quasars, \citet{Kulkarni2022} have shown that nearly all of them are either DLAs or sub-DLAs.

The alternate approach to identifying the GOTOQs is to search for associated nebular emission lines (e.g., \OII) in the quasar spectra at the known absorption redshift. Starting from the \MgII\ $\rm{\lambda\lambda\, 2796, 2803}$ absorption doublet present on the background quasar spectra, if we search for associated \OII\ $\rm{\lambda\lambda\, 3727, 3729}$ emission which is ubiquitous in all star-forming galaxies, it will allow us to study the disk-halo interface of galaxies over the redshift range $0.35 \leqslant z \leqslant 1.5$. \citet{Joshi2018} have identified 198 GOTOQs associated with the \MgII\ absorbers (we provide more details in Section~\ref{sec:data}). The host galaxy properties (in terms of impact parameter distribution, stellar mass, star formation rate, etc.) are not studied in detail till now. The availability of nearly uniform high-quality imaging data from 
the Dark Energy Spectroscopic Instrument (DESI) Legacy Imaging Survey \citep{Dey2015, Dey2019DESI} enables us to undertake such a study. This forms the main focus of this paper.

This paper is organized as follows. In section \ref{sec:data}, we describe our sample and the corresponding spectroscopic and photometric data used in this work. In {Section \ref{sec:galaxy_properties}, we measure  the properties of the individual host galaxies where it is possible to  decompose the galaxy image from the quasar image. We also explore the correlations between the galaxy properties and properties of the absorption lines detected in the quasar spectra. In section \ref{sec:stacking}, we use the stacked images and spectra to derive the average properties of the host galaxies of GOTOQs.  Our findings on the nature of host galaxies of GOTOQs are summarised in section \ref{sec:summary}.} For this work, we assume a flat $\Lambda$-CDM cosmology $H_0 = \rm{70\, km\, s^{-1}\, Mpc^{-1}}$ and $\Omega_m = 0.3$.

\section{Sample \& Data}
\label{sec:data}
For this work, we consider the sample of GOTOQs associated with the known \MgII\ absorbers compiled by \citet{Joshi2017}. In brief, \citet{Joshi2017} inspected all the quasar spectra showing intervening \MgII\ absorption (with $\rm{W_{2796}\, \geqslant\, 0.1\text{\AA}}$) listed in the SDSS Fe~{\sc ii}/\MgII\ metal absorber catalog of \citet{Zhu2013} and searched for the associated \OII\ $\lambda\lambda$ 3727, 3729 nebular emission lines detected with more than $\rm{4\sigma}$ significance. This has resulted in a total of 185 GOTOQs in the redshift interval 0.35 $\leqslant z_{abs} \leqslant$ 1.05. It is known that for a small fraction of high-$z$ galaxies, the \OIII\ $\lambda\lambda$ 4960, 5008 nebular emissions can be stronger than the \OII\ $\lambda\lambda$ 3727, 3729. To account for such galaxy populations, \citet{Joshi2017} searched for the \OIII\ emission doublets, detected with more than $3\sigma$ significance, associated with the \MgII\ absorption. This has resulted in 13 more GOTOQs based on \OIII\ emission for which the \OII\ nebular emissions are detected at $<4\sigma$ significance. Therefore, their sample contains a total of 198 GOTOQs. Among these GOTOQs, 67 were detected in SDSS-DR7 (that uses a fiber with a projected angular diameter of 3 arcsec), 117 were detected in SDSS-DR-12 (that uses a fiber with a projected angular diameter of 2 arcsec), and the rest of the 14 GOTOQs were observed both in SDSS-DR7 and SDSS-DR12.

Deep images of every quasar in this GOTOQ sample are available from the DESI Legacy Imaging Survey \citep{Dey2015, Dey2019DESI}. We obtain images in all the available $grz$ photometric bands. The DESI Legacy Imaging Survey is known to be complete up to an apparent $r$ band magnitude of 23.6 mag. Assuming $M_r^\star = -21.74$ mag \citep{Karademir2022} and the k-correction to be negligible, this magnitude limit at the median redshift (i.e., \zabs $\sim$ 0.665) of our GOTOQ sample  corresponds to $\ge 0.12M_r^\star$ galaxies. However, note that the detection sensitivity may not always reach this magnitude limit due to the presence of a bright quasar close to the foreground galaxy. We use the images from DESI Legacy Imaging Survey to study the nature of GOTOQs by direct decomposition of quasar and galaxy contributions and also using image stacking techniques. We also use all the SDSS spectra of quasars in the GOTOQ sample to investigate the line of sight reddening, mean stellar and photospheric absorption, and nebular emission properties of the GOTOQs using spectral stacking techniques. 

\section{Foreground Galaxy Properties}
\label{sec:galaxy_properties}
By construction, the quasar-galaxy pairs in the GOTOQ sample have exceptionally small impact parameters. Since the foreground galaxies are typically much fainter than the background quasars, they are either completely outshined by the background quasar or sometimes detected as an extension around the quasar in broadband photometry. Whether or not the foreground galaxy would be detected as a photometric extension to the quasar depends not only on the brightness of the background quasar and the foreground galaxy but also on the impact parameter and the orientation of the foreground galaxy with respect to the quasar sightline as well. Therefore, we individually inspected all the GOTOQ images from the DESI Legacy Imaging Survey and searched for the presence of photometric extensions around the background quasars beyond the point spread function (PSF).

\subsection{Identifying the foreground galaxies}

\begin{figure*}
    \centering
    \includegraphics[viewport=5 5 890 300,width=0.97\textwidth,clip=true]{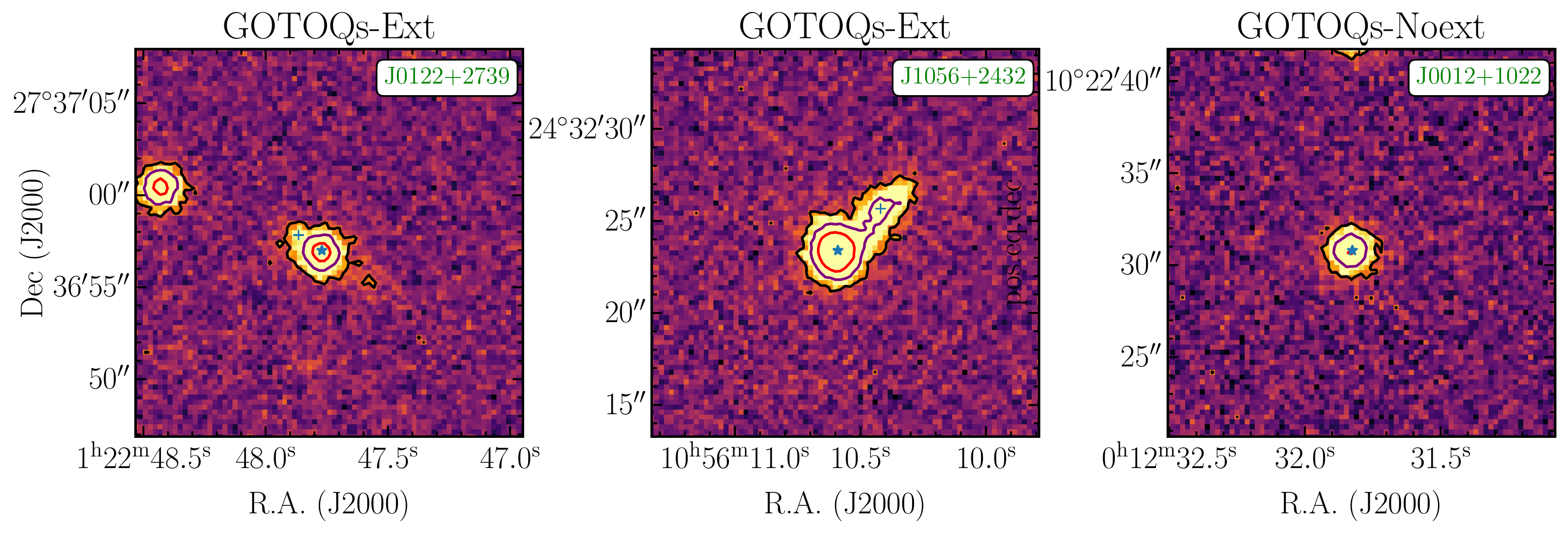}
    \caption{{\it Left and Middle panels: } Illustations of GOTOQs showing the photometric extensions. The three panels show the DESI r-band images of three GOTOQ in our sample. {\it Right panel:} An example of quasar with a GOTOQ along the line of sight without showing any extension. For each panel, the quasar is marked with a blue `$\star$', and the contours correspond to the $3\sigma,\, 10\sigma,\, $ and $30\sigma$ levels above the mean background. The centroids of the photometric extensions are marked with a blue `+'. These were also identified as unique photometric source in the DESI Legacy
Image Survey catalogs.}
    \label{fig:vis_inspec}
\end{figure*}

\label{sec:host_galaxies}
\begin{figure*}
    \centering
    \includegraphics[viewport=5 10 950 275,width=\textwidth,clip=true]{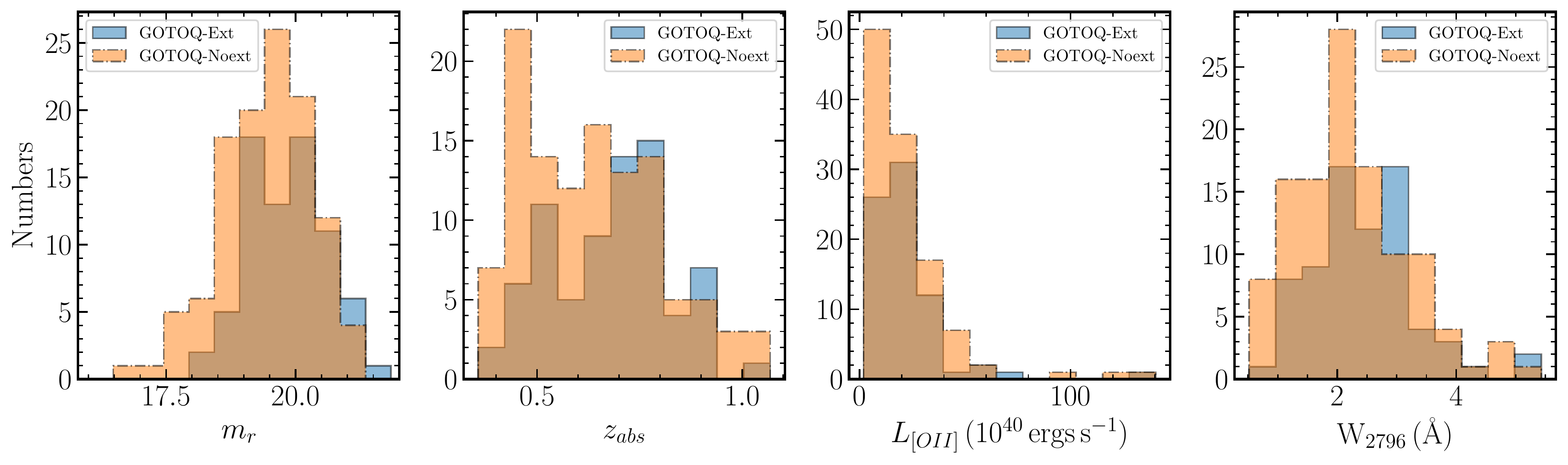}
    \caption{Comparison between systems with the photometric extensions (GOTOQ-Ext; blue histogram) and the systems without them (GOTOQ-Noext; orange histogram) based on various spectroscopic and photometric properties. From left to right, each panel shows the distributions of apparent r-band magnitude of the background quasar,  \MgII\ absorption redshift,  \OII\ line luminosity, and  $\rm{W_{2796}}$.
    }
    \label{fig:compare_ext_noext}
\end{figure*}

Purely based on the visual inspections of all the 198 GOTOQs in our sample, we have identified a total of 98 GOTOQs where another source is visible as a photometric extension around the background quasar. The method of identifying photometric extensions around the quasar based on visual inspections is illustrated in Figure \ref{fig:vis_inspec}. However, to make sure the identified photometric extensions are due to the foreground galaxies detected in emission on top of the quasar spectra and not due to chance coincidence of unrelated galaxies or stars, we consider only those cases for which the photometric redshift of the extensions are consistent within $2\sigma$ of the nebular emission redshifts. This brings down the total number of GOTOQs with the foreground galaxy detected as a broadband photometric extension to 84. The photometric redshifts of the sources associated with the photometric extensions are also obtained from the DESI Legacy Image Survey. The photometric redshifts are computed using the random forest algorithm. Details of the photo-z training and performance can be found in \citet{Zhou2021}. They found a typical redshift uncertainty to be of the order 0.062 for objects with r-band magnitude brighter than the 23rd mag. The redshift uncertainty is larger than this for fainter sources.

Using the South African Large Telescope (SALT) spectroscopic follow-up of a \OII\ line luminosity limited sub-sample consisting of 16 GOTOQs, we find that the photometric extensions with consistent photometric redshift are indeed due to foreground galaxies detected in emission in the quasar spectra (Guha et al., in preparation). Therefore, as in \citet{Straka_2015}, we consider the above-mentioned 84 galaxies to be foreground galaxies responsible for the \OII\ emission and the \MgII\ absorption detected in the spectra of background quasars. From hereon, these sample GOTOQs for which the foreground galaxies are detected as broadband photometric extensions are referred to as `GOTOQ-Ext'.  

Among these 84 GOTOQs, 25 were originally observed in SDSS-DR7, 56 were observed in SDSS-DR12, and the rest of the three were observed in both SDSS-DR7 and SDSS-DR12. For ten out of these 84 systems, the background quasars and the foreground galaxies are well separated in the sky plane, and the overlap between them is the bare minimum. Out of these ten GOTOQs, six are observed in SDSS-DR12, while the rest of the four are observed in SDSS-DR7. Given the typical SDSS seeing of 1.3$^{\prime\prime}$, some fluxes from these galaxies may have leaked into the SDSS fiber used to observe the background quasar, thereby making these systems appear as GOTOQs. However, to be on the conservative side, we exclude them from our sub-sample of the GOTOQ-Ext as the true host galaxy may just sit on top of the quasar without producing any detectable photometric extensions. The details of these 84 systems with detected photometric extensions are listed in the Appendix in Table \ref{tab:extended_GOTOQ}. The ten systems excluded are marked by an asterisk. Therefore, we have 74 systems in the sub-sample of ``GOTOQ-Ext".

For the remaining 114 GOTOQs (called the "GOTOQ-Noext" sample), we do not find any significant photometric extensions around the quasar with consistent photometric redshifts. This fraction is much higher in our sample \citep[i.e. $\sim$ 58\% compared to 31\% found for the sample of][]{Straka_2015}. The non-detection of any photometric extension in these 114 systems could stem either from one or the combination of the following reasons. The background quasars are relatively bright, the foreground galaxies are comparatively faint, and, the impact parameters are relatively small. 
\begin{table}
  \caption{Results of KS-test between the GOTOQ-Ext and GOTOQ-Noext sub-samples based on various photometric and spectroscopic properties.}
    \centering
    \begin{tabular}{lcccc}
    \hline
         Result & $m_r$ & \zabs & $L_{[O\, \textsc{ii}]}$ & $W_{2796}$ \\
         \hline
         $p$ & 0.023 & 0.059 & 0.125 & 0.198\\
         $D$ & 0.219 & 0.193 & 0.171 & 0.156 \\
         \hline
    \end{tabular}
    \label{tab:ks_test}
\end{table}

The Kolmogorov–Smirnov (KS) test between these two sub-samples (GOTOQ-Ext and GOTOQ-Noext) based on various properties are summarised in Table \ref{tab:ks_test}. The distributions are shown in Figure~\ref{fig:compare_ext_noext}. The KS-tests yield $p$ values of more than 0.05 for absorption redshift ($z_{abs}$), \OII\ line luminosity, and the rest equivalent width of \MgII$\lambda$2796 ($W_{2796}$) implying that the difference between these two sub-samples are statistically insignificant as far as these properties are concerned. 

In the histogram  plot of the r-band magnitude of the quasars, it is apparent that when the background quasars are relatively bright (low $m_r$), the fraction of GOTOQs having photometric extensions drops. Consequently, the KS-test yields a $p$ value of 0.023. Other than the brightness of the background quasar, there are two additional factors (the brightness of the host galaxies and their impact parameters) that are important for the detection of the foreground galaxies as photometric extensions. In section \ref{sec:stacking}, using the method of image stacking, we will explore the role of these two factors.

\begin{figure}
    \centering
    \includegraphics[viewport=5 5 350 330,width=0.5\textwidth,clip=true]{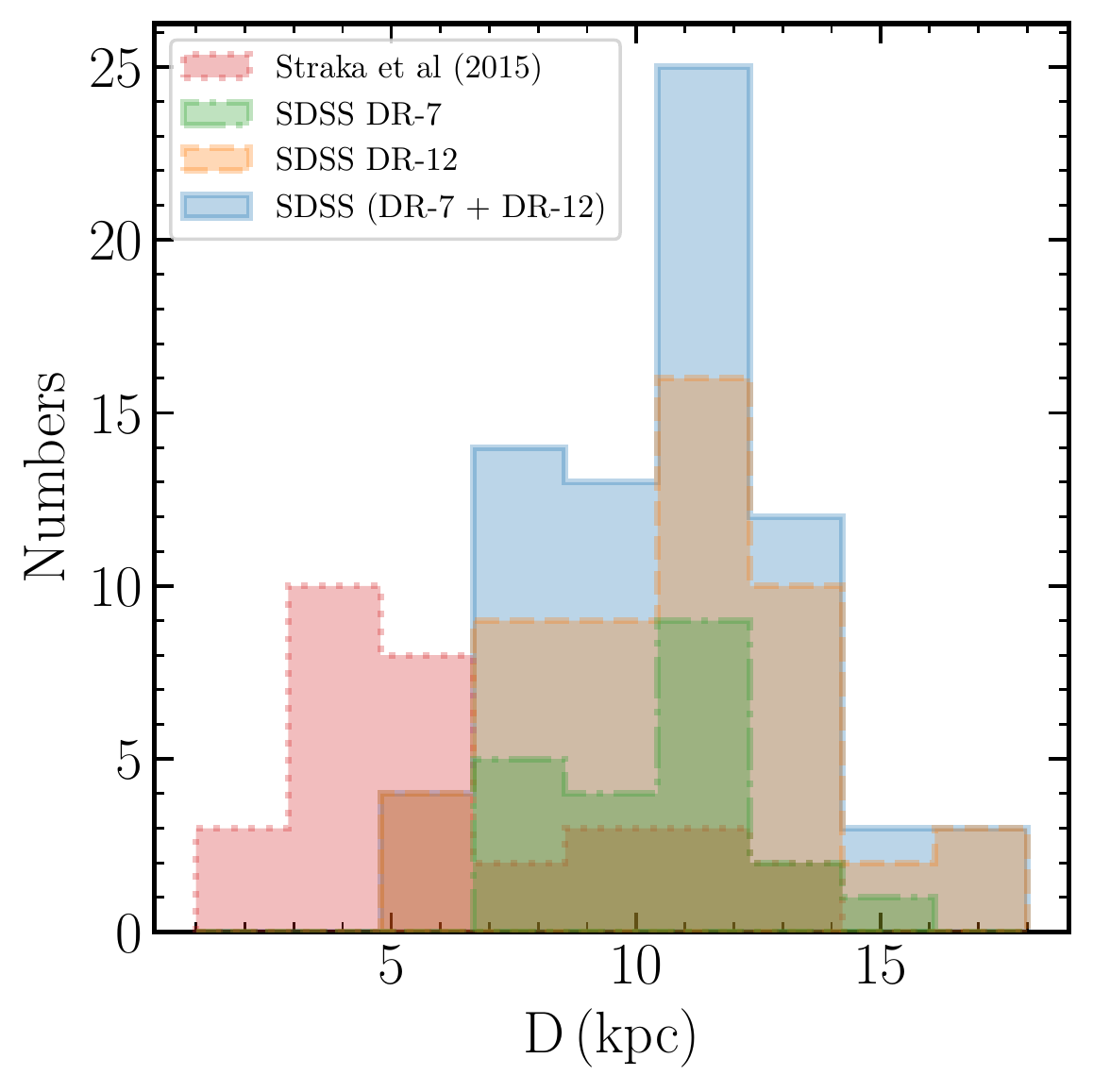}
    \caption{Comparison of the impact parameter distributions of low-$z$ GOTOQs identified by \citet[][]{Straka_2015} (red histogram) with that of 
    objects in our ``GOTOQ-Ext" sample (blue histogram). The green and the orange histograms respectively show the impact parameter distributions of the objects  in the ``GOTOQ-Ext" sample observed in SDSS DR-7 and SDSS DR-12.}
    \label{fig:impact_compare}
\end{figure}

\begin{table}
    \centering
     \caption{Results of Spearman Rank correlation.}
   \begin{tabular}{cccr}
    \hline
         Property 1& Property  2& $r_S$ & $p$ value \\
         \hline
          D& $W_{2796}$ & -0.097 & 0.413 \\
           & $W_{2852}$ & -0.074 & 0.546 \\
           & $\mathcal{R}$& -0.140 & 0.251 \\  
           & $V_{\rm{Mg\textsc{ii} - [O\textsc{ii}]}}$ & 0.029 & 0.801\\
           & $z$ & 0.282 & 0.015 \\
           & $L_{[O\,\textsc{ii}]}$ & 0.157 & 0.182 \\
            & $m_r^{\rm{qso}}$& -0.032 & 0.784 \\
        \\
         $M_B$ & D &  -0.464  & $3.123\times10^{-5}$  \\
               & $z_{gal}$  &  -0.183  & 0.118\\
               & $W_{2796}$ &  -0.234  & 0.045\\
\hline
    \end{tabular}
    \label{tab:correlation}
\end{table}

\subsection{Individual measurements of impact parameters for the GOTOQ-Ext}
\begin{figure*}
    \centering
    \includegraphics[viewport=5 5 623 620,width=0.9\textwidth,clip=true]{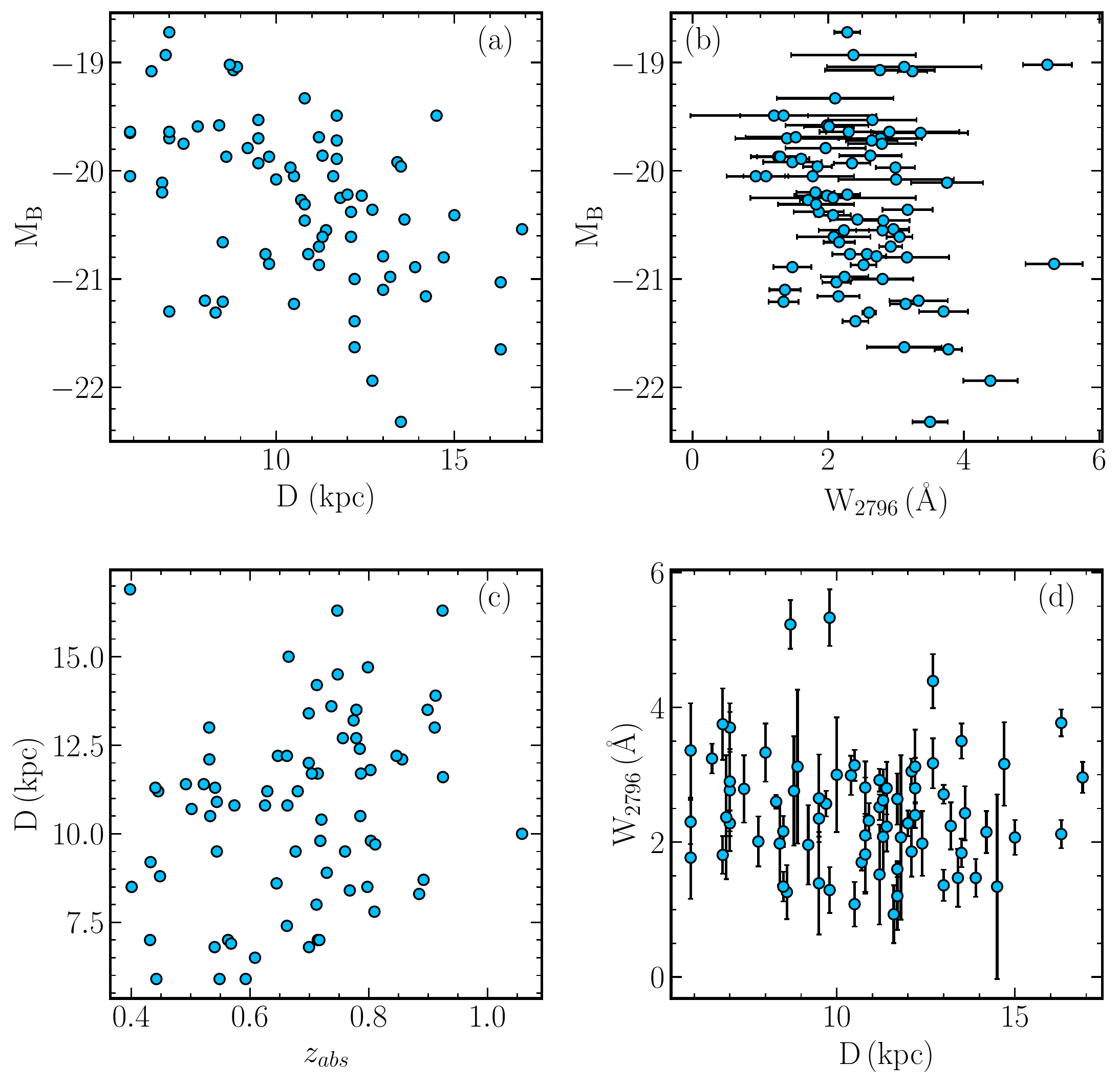}
    \caption{The plot showing significant correlations / anti-correlations found between different parameters in the sub-sample of ``GOTOQ-Ext". Panel (a) shows the absolute rest frame B-band magnitude of foreground galaxies versus the impact parameter. Panel (b) shows the absolute rest frame B-band magnitude of foreground galaxies versus the rest frame equivalent width of \MgII\ $\lambda$ 2796 line. Panel (c) shows the impact parameters versus the absorption redshift, while Panel (d) shows $W_{2796}$ against the impact parameter.}
    \label{fig:galaxy_correlations}
\end{figure*}

We obtain the impact parameters of the host galaxies for the objects in the ``GOTOQ-Ext" sample using the decomposed locations of quasar and galaxy in the DESI Tractor Catalog\footnote{https://www.legacysurvey.org/dr9/description/\#tractor-catalogs-1} \citep[see section 8 of][]{Dey2019DESI}. The projected physical separation between centroids of quasar and galaxy, obtained for the cosmological parameters used in this work, is the impact parameter of the galaxy. In Figure \ref{fig:impact_compare}, we compare the impact parameter distributions between objects in the ``GOTOQ-Ext" sample with those identified based on the nebular emission line searches at relatively low-redshifts \citep[][hereafter refer to as S15 sample]{Straka_2015}.
In the case of the ``GOTOQ-Ext" sample, we use the \MgII\ absorption redshift to be the redshift of the identified galaxy.

The red histogram in this figure corresponds to the impact parameter distribution of the S15 sample and the blue histogram corresponds to the same for objects in the ``GOTOQ-Ext" sub-sample. The green and the orange histograms respectively show the impact parameter distributions of the objects in the ``GOTOQ-Ext" sample observed in SDSS DR-7 and SDSS DR-12. The impact parameter of galaxies detected in S15 ranges from 0.37 kpc and 12.68 kpc with a median value of 4.83 kpc. Whereas the same in the case of our ``GOTOQ-Ext" sample ranges from 5.9 kpc to 16.9 kpc with a median value of 10.85 kpc. From the impact parameter distributions, it is apparent that compared to our GOTOQ-Ext  sample, the impact parameters of the objects in the S15 sample are smaller. A two-sample KS test yields an extremely low p-value of $8\times10^{-11}$, confirming that the impact parameter distributions for these two samples are very different. 

This result is mainly due to selection bias. The S15
sample, which contains low-redshift galaxies, is biased against detecting galaxies at large impact parameters as the projected length scale for a fiber of a given radius will be lesser at these redshifts compared to redshifts of objects in our sample. It is also well documented that at a given redshift detection of nebular \OII\ emission from GOTOQ is biased by the size of the fiber used. To see if the impact parameter distribution (that we obtain using the DESI Legacy Imaging Survey data) is biased by fiber size used in the spectroscopy to identify GOTOQs, we performed a two-sample KS test between the impact parameters of GOTOQs observed in SDSS-DR7 and SDSS-DR12. A high p-value of 0.42 obtained implies that the impact parameter distributions in our sample obtained here are not biased by the fiber size effects. Therefore, in what follows we do not treat GOTOQs detected from SDSS-DR7 and SDSS-DR12 differently when studying the galaxy properties.

Other than providing coordinates, the Tractor catalog also provides galaxy fluxes in different available photometric bands. We compute the apparent r-band magnitudes using the measured r-band fluxes and then convert this to the absolute rest frame B-band magnitude using the distance modulus of the absorbers and assuming the average SED fitted spectrum (see Section \ref{sec:stacking}). The measured rest frame absolute B-band magnitude ranges from -22.32 to -18.72 mag. With $M_B^\star$ = -21.53 \citep{Faber2007}, this corresponds to a rest frame B-band luminosity range of 0.075$L_B^\star$ to 2.07$L_B^\star$.

\subsection{Correlation between absorption properties and galaxy properties for the GOTOQ-Ext}

\begin{figure*}
    \centering
    \includegraphics[viewport=10 7 550 270,width=0.9\textwidth,clip=true]
    {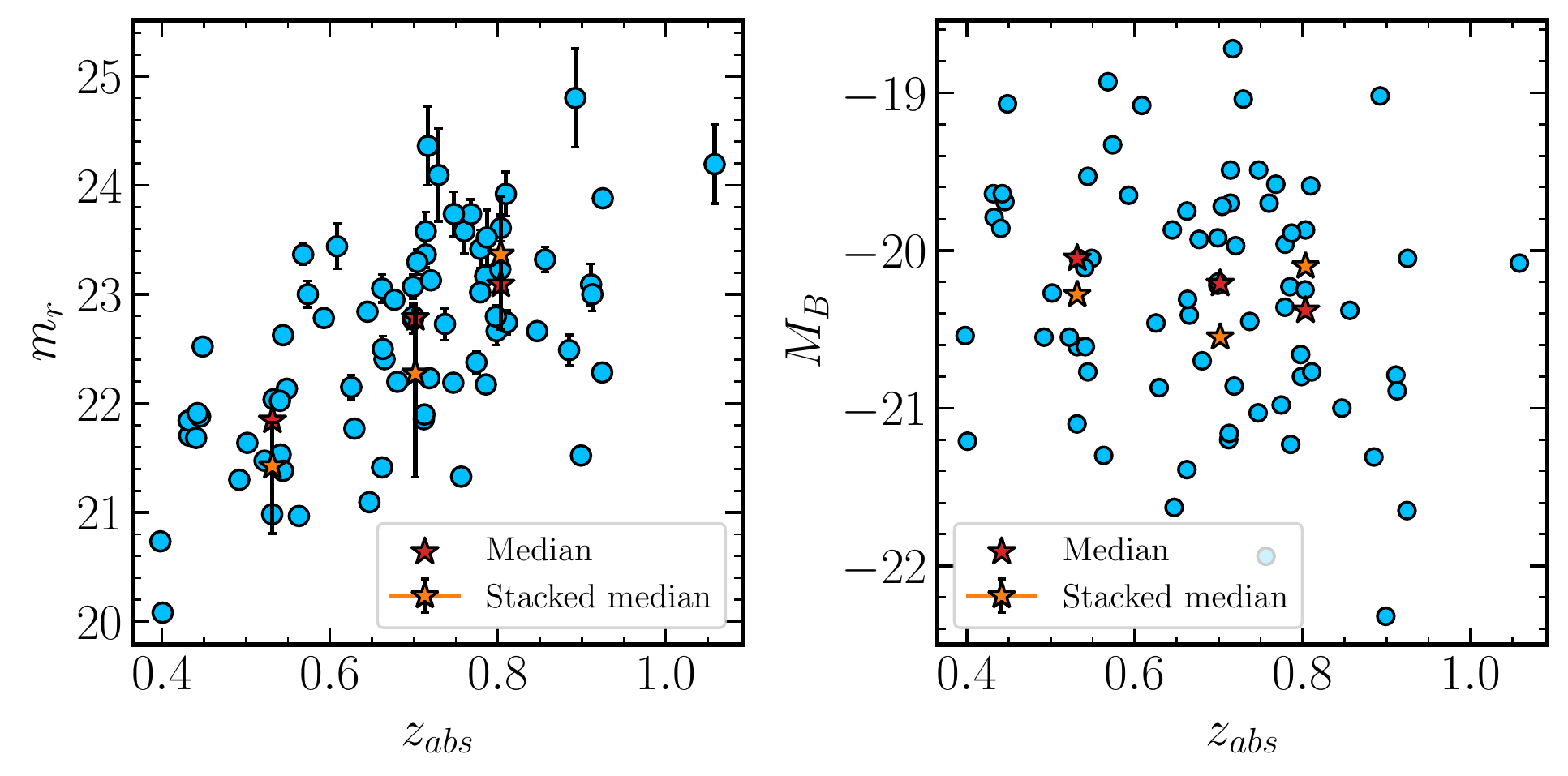}
    \caption{\textit{Left panel} shows the r-band apparent magnitude of the foreground galaxies obtained from the DESI Legacy Survey Tractor catalog against the absorption  redshifts. The red stars correspond to the median values of the apparent r-band magnitudes in the three redshift ranges
    considered here. The orange stars correspond to the measured r-band magnitudes in each of these bins obtained from the image stacking (see discussions in section~\ref{sec:stacking}). The error bar shown is the 16th to 84th percentile range obtained using bootstrapping. \textit{Right panel} shows the same but for the rest frame B-band absolute magnitude calculated assuming the average SED spectrum.}
    \label{fig:zabs_vs_rmag}
\end{figure*}

In this section, we investigate possible correlations between the foreground galaxy properties (mainly impact parameter and B-band absolute magnitude, $M_B$) and properties derived using absorption lines in the case of ``GOTOQ-Ext" sub-sample. Results of Spearman correlation are summarized in Table~\ref{tab:correlation}. The first two columns in this table list the properties compared and the last two columns give the correlation coefficient and p-value.

First, we examine the correlations between different parameters  and the impact parameter. It is evident from the last column in Table~\ref{tab:correlation} that there is no statistically significant correlation between the impact parameter and equivalent width of Mg~{\sc ii}, Mg~{\sc i}, and Fe~{\sc ii} to Mg~{\sc ii} equivalent width ratio (i.e $\mathcal{R}$). In particular, the lack of correlation between $W_{2796}$ and D  in our sample suggests a possible flattening in the well-known anti-correlation between these quantities at small D. This can be seen directly from the panel (d) of Figure \ref{fig:galaxy_correlations}. We discuss this in detail in section~\ref{sec:impact_param}. The lack of any statistically significant difference in the $W_{2796}$ distribution between the ``GOTOQ-Ext" and ``GOTOQ-Noext" sub-samples (Table~\ref{tab:ks_test} and Figure~\ref{fig:compare_ext_noext}) are also consistent with the flattening of the distribution at low D.

We do not find any correlation between the velocity difference measured between \MgII\ absorption redshift and \OII\ emission redshift (denoted by 
$V_{\rm{Mg\textsc{ii} - [O\textsc{ii}]}}$) and impact parameters.
We do not also find any correlation between \OII\ luminosity measured from the fiber spectra and impact parameter. Naively one expects the \OII\ luminosity to be lower when the impact parameter is larger as more and more light will go through the fiber when D is small. However, this does not seem to strongly affect our sample.

Purely based on observational effects, we expect a possible anti-correlation between quasar magnitude and impact parameter. This is because when the quasar is bright, our ability to detect a galaxy at low impact parameters will be impaired. We do not see a statistically significant correlation even in this case.  Only a marginally significant ($p$-value of 0.015) correlation is seen between $z_{abs}$ and D. The impact parameters against the absorption redshifts for the ``GOTOQ-Ext" sub-sample are plotted in the panel (c) of Figure~\ref{fig:galaxy_correlations}. As discussed before, this could be related to the physical distance probed for a given angular scale being higher at higher redshifts.

Note, host galaxies of Mg~{\sc ii} absorbers identified in the \textsc{magiicat} survey \citep{Nielsen_2013} tend to show  anti-correlation between impact parameter and absolute B-band magnitude \citep[][]{Guha2022}. As can be seen from the panel (a) of Figure \ref{fig:galaxy_correlations}, there is a strong anti-correlation between the impact parameter and $M_B$ with a correlation coefficient of $-0.464$ and a p-value of $3.123\times10^{-5}$ in our sample as well. We also find, a possible anti-correlation (with a correlation coefficient of $-0.234$ and $p$-value of 0.045) between $M_B$ and W$_{2796}$. This is also shown in panel (b) of Figure~\ref{fig:galaxy_correlations}. This suggests that the brighter galaxies are typically associated with stronger \MgII\ absorption, albeit with a large scatter. If the absorbing galaxies in the ``GOTOQ-Noext" sample are indeed faint, then the above-mentioned correlation may imply a statistically low W$_{2796}$ for this sub-sample. However, as discussed above, the W$_{2796}$ distributions for the two sub-samples are statistically indistinguishable. This implies that the main reason for the non-detection of photometric extension in ``GOTOQ-Noext" is related to the impact parameter being low and/or the background quasar being relatively bright. Below we address this in detail using stacked images.

In Figure~\ref{fig:zabs_vs_rmag}, we plot the apparent ($m_r$ in the left panel) and the rest frame B-band absolute ($M_B$ in the right panel) magnitudes of galaxies against the absorption redshift. Even though there is a strong correlation seen between $m_r$ and $z$, it is evident from this figure and Table~\ref{tab:correlation} that there is no clear correlation between $M_B$ with \zabs over the small redshift range considered here. The Spearman rank correlation coefficient is $-0.183$ with a p-value of 0.118. 

\subsection{$W_{2796}$ vs D correlation for galaxies in the GOTOQ-Ext sample}
\label{sec:impact_param}

\begin{figure*}
    \centering
    \includegraphics[viewport=10 10 850 330,width=\textwidth,clip=true]{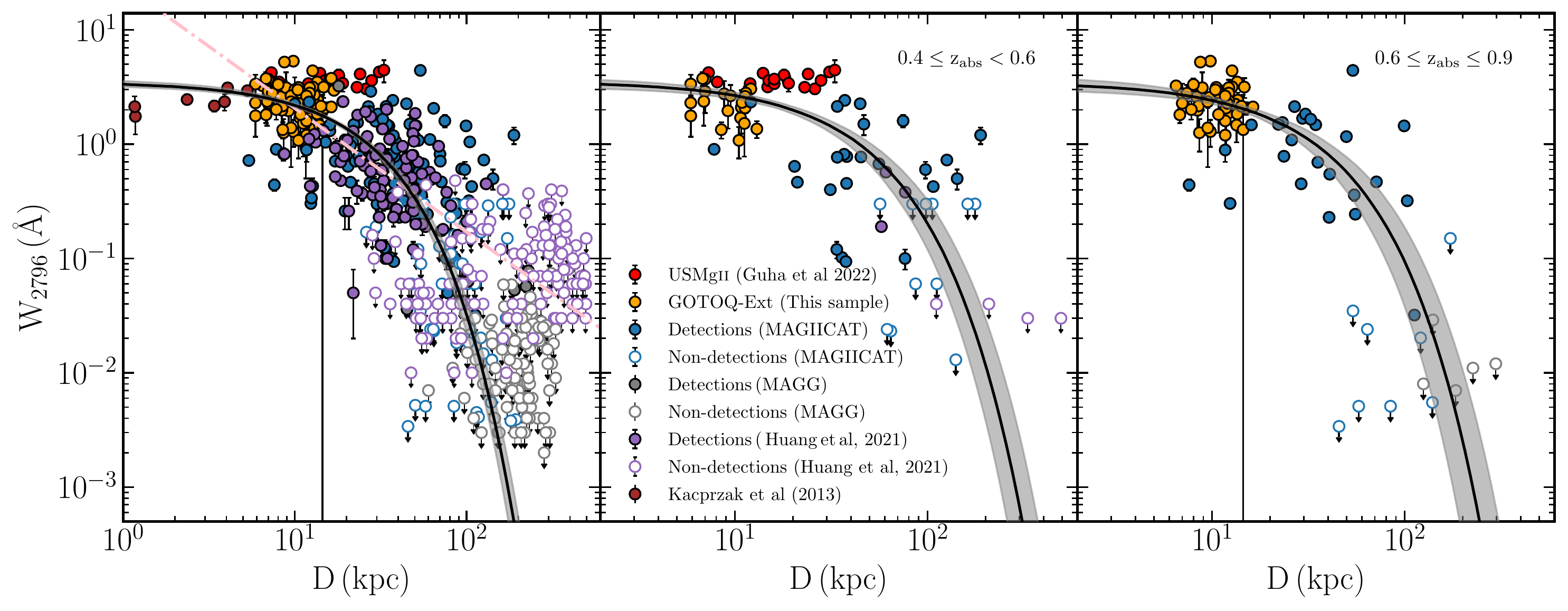}
    \caption{The \emph{left panel} shows the impact parameter (D) versus the $\rm{W_{2796}}$ anti-correlation over the full redshift range for the isolated galaxies. Orange points are for the objects in the ``GOTOQ-Ext" sub-sample. The red, blue, violet, gray, and brown points are taken from \usmg\ survey \citep{Guha2022}, MAGIICAT survey \citep{Nielsen_2013}, \citet{Huang2021}, MAGG survey\citep{Dutta2020}, and \citet{Kacprzak_2013} respectively. Note that in the MAGG survey, almost two-thirds of the \MgII\ absorption systems are associated with more than one galaxies \citep{Dutta2020, Dutta2021}. We consider only the \MgII\ absorption systems associated with isolated galaxies. The solid black line corresponds to the best-fit log-linear model, and the shaded region corresponds to the $1\sigma$ errors associated with it. The GOTOQs typically follow the average anti-correlation. The \emph{middle } and \emph{right panel} show the same, but only for the redshift ranges, $0.4  \leqslant z_{abs} < 0.6$, and $0.6 \leqslant z_{abs} \leqslant 0.9$ respectively.}
    \label{fig:w_vs_rho}
\end{figure*}

{In the left panel of Figure \ref{fig:w_vs_rho}, we plot the impact parameters obtained for objects in the ``GOTOQ-Ext" sub-sample against $\rm{W_{2796}}$. This plot also shows the same for the host galaxies of \MgII\ absorption systems studied in the literature \citep{Kacprzak_2013, Nielsen_2013, Dutta2020,  Huang2021, Guha2022}. As expected, our measurements have substantially increased the number of systems in the impact parameter range 5.9 to 16.9 kpc and provide stringent constraints on the $\rm{W_{2796}}$ distribution at low impact parameters. Excluding the ``GOTOQ-Ext" sub-sample discussed here and the \usmg\ sample of \citet{Guha2022}, only 35 \MgII\ host galaxies are identified in the literature sample with $D\le17$ kpc over the entire redshift range. However, these systems are primarily from the low redshifts. There are only 10 and 4 systems present in the redshift range $0.4 \leqslant z_{abs} < 0.6$ and $0.6 \leqslant z_{abs} \leqslant 0.9$ respectively. The inclusion of 74 systems from the ``GOTOQ-Ext" sub-sample substantially increases the total number of \MgII\ absorption systems in the low impact parameter ($D < 17$ kpc) range. Our sample, therefore, is important in identifying the \MgII\ equivalent width at which the $\rm{W_{2796}}$ vs. D relationship flattens, the characteristic impact parameter, and their redshift evolution.

The solid black line  in Figure~\ref{fig:w_vs_rho} shows the maximum likelihood fit to the data assuming a log-linear function of the form $\log\,W_{2796} = \alpha D\, (\rm{kpc}) + \beta $. To appropriately take into account the upper limits in the measurements of $\rm{W_{2796}}$, we use the maximum likelihood method following the standard approach given in the literature \citep{Chen_2010, Rubin2014, Dutta2020, Guha2022}. Note the above expression is identical to ${W_{2796} (D)  = W_{2796} (D=0) exp (-D/h)}$, with $W_{2796} (D=0) = 10^{\beta}$ and characteristic impact parameter scale $h = 1/(2.303\times\alpha)$ kpc.

The best fit parameter for different sub-samples are summarized in Table~
\ref{tab:fitted_params}. The gray region around the solid line in Figure~\ref{fig:w_vs_rho} corresponds to $1\sigma$ uncertainty to the fit.
For the full dataset (without any redshift cut) the best-fitted parameters obtained are $\alpha = -0.020\pm0.002$, $\beta = 0.537\pm0.025$ with an intrinsic scatter of $\sigma = 0.884\pm0.036$.  This table also suggests that the exclusion of \usmg\ absorbers from the sample has very little effect on the derived parameters. The dot-dashed pink line corresponds to the power-law fit of the form $\log\, W_{2796} = \alpha\, \log\, D\, (\rm{kpc}) + \beta $ with $\alpha$ = 24.9$\pm$3.1, $\beta$=1.082$\pm$0.053, and $\sigma$=0.979$\pm$0.043. As indicated in previous studies, the power law fit overestimates the $\rm{W_{2796}}$ at low $\rm{D}$ and has a larger scatter at large D. Therefore, in what follows, we mainly use the log-linear fits.

Our best fit suggests that $\rm{W_{2796} (D=0)} = 3.44\pm 0.20$\AA\ and $h = 21.6_{-1.97}^{+2.41}$ kpc. If we do not consider the GOTOQs from our sample, the best-fitted parameters are $\alpha = -0.019\pm0.002$, $\beta = 0.464\pm0.039$ with an intrinsic scatter of $\sigma = 0.914\pm0.042$ \citep{Guha2022}. This suggests that $\rm{W_{2796}} (D=0) = 2.91\pm 0.26$\AA\ and $h = 22.85^{+2.69}_{-2.17}$ kpc.
The value of $\rm{W_{2796}} (D=0)$ we find here are higher than 1.87$\pm$0.47\AA\ found 
by \citet{Kacprzak_2013} and 0.89$^{+1.45}_{-0.53}$\AA\  by \citet{Dutta2020}. It is evident from Figure~\ref{fig:w_vs_rho} that three out of 7 of low-$z$ points, that defined the $\rm{W_{2796}} (D=0)$ in all previous studies,  are below our fit by more than 3$\sigma$ level. 
These differences noted could be related to either of (i) small number statistics at $D<6$ kpc, (ii) redshift evolution (our data points are predominantly at $z>0.36$ compared to $z\sim0.1$ for the low-D data points from \citet{Kacprzak_2013}) and (iii) intrinsic deviation from the smooth fit at low-D due to different feedback processes affecting the gas distribution. Therefore, it is important to increase the number of measurements at $D<5$ kpc. In particular, identifying host galaxies in the case of the ``GOTOQ-Noext" sub-sample using either space-based imaging or ground-based adaptive optics (AO) supported imaging may help in this regard.

\begin{table}
    \centering
    \caption{The best fitted parameters for log-linear characterization of the $W_{2796} - D $ anti-correlation shown in Figure \ref{fig:w_vs_rho}. Redshift ranges marked with $\star$ are for the fit without the \usmg\ systems.}
    \begin{tabular}{lccc}
    \hline
         Redshift & $\alpha$ & $\beta$ & $\sigma$\\
         \hline
         0.09 - 1.49 & -0.020$\pm$0.002 & 0.537$\pm$0.025 & 0.884$\pm$0.036 \\
         0.09 - 1.49$^\star$ & -0.022$\pm$0.002 & 0.526$\pm$0.028 & 0.792$\pm$0.035\\
         0.4 - 0.6   & -0.012$\pm$0.003 & 0.531$\pm$0.046 & 1.090$\pm$0.100 \\
         0.4 - 0.6$^\star$   & -0.019$\pm$0.005 & 0.514$\pm$0.065 & 0.782$\pm$0.091 \\
         0.6 - 0.9   & -0.017$\pm$0.004 & 0.547$\pm$0.053 & 1.007$\pm$0.098 \\
         \hline
    \end{tabular}
    \label{tab:fitted_params}
\end{table}

Next, we investigate whether is there any redshift evolution in the $W_{2796}$ vs D relationship by considering two sub-samples over the redshift range $0.4 \le$\zabs$\le 0.9$.  In the middle and the left panels of  Figure~\ref{fig:w_vs_rho}, we plot the same for the redshift ranges $0.4 \leqslant z_{abs} < 0.6$, and, $0.6 \leqslant z_{abs} \leqslant 0.9$ respectively. Note in these redshift ranges our ``GOTOQ-Ext" sub-sample contributes about 36\% and 65\% of the total sample.
The fit parameters are summarized in Table~\ref{tab:fitted_params}. It is clear from the table that the value of $\beta$ (and hence $W_{2796}(D=0)$) and $\alpha$ (and hence h) are consistent between the two sub-samples. Thus, within the uncertainties in the derived parameters, we do not see any statistically significant redshift evolution. It is also evident from the table that this result does not change even when we do not consider the data points for \usmg\ sample for $0.4 \leqslant z_{abs} < 0.6$.
Lack of redshift evolution in the $W_{2796}$ $vs.$ D relationship was already discussed in the literature \citep[see for example, section 3.6 of][]{Dutta2020}.
}

\section{Results from the stacking analysis}
\label{sec:stacking}
As we discussed before, we do not have a clear identification of host galaxies of GOTOQs in 58\% of the cases. Here, we use image and spectral stacking techniques to draw some inferences on the nature of the host galaxies of these GOTOQs. 

\subsection{Stacking of quasar images}

\begin{figure*}
    \centering
    \includegraphics[viewport=5 5 750 380,width=0.7\textwidth,clip=true]{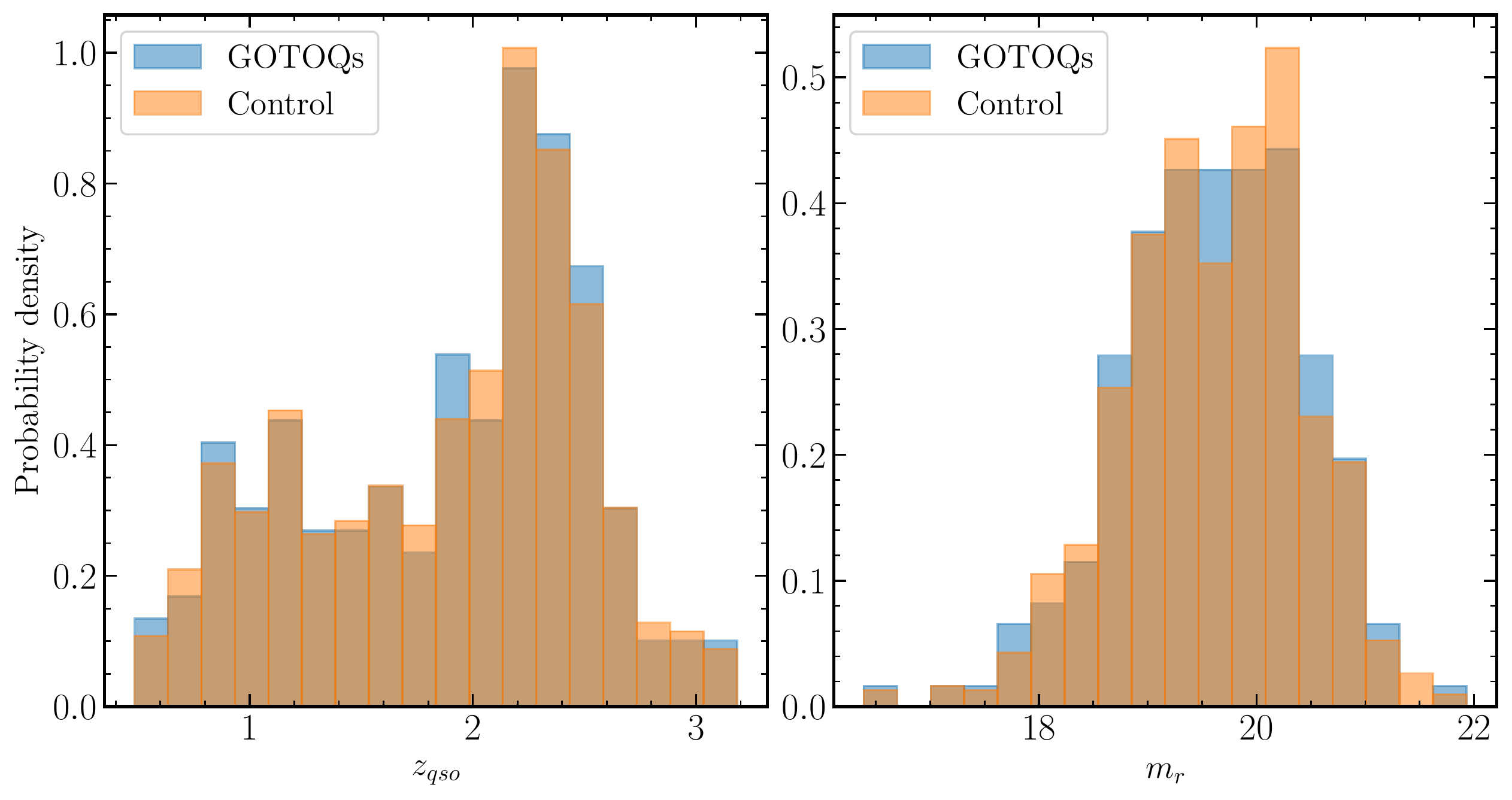}
    \caption{Comparison of the distributions of emission redshift (left panel) and r-band magnitude (right panel) of quasars in the GOTOQ sample (blue) and in the control sample (orange). As designed, these distributions are consistent with being drawn from the same parent population.
    }
    \label{fig:compare_samples}
\end{figure*}

We obtained the multiband ($g$, $r$, and $z$ bands) deep photometric images of all the 198 GOTOQs from the DESI legacy imaging Survey. For each GOTOQ, we have downloaded a 256$\times$256 pixels cutout of the image centered around the quasars. These are flux-calibrated and continuum-subtracted images. The width of each pixel in these images corresponds to the angular scale of 0.27 arcsec. Before we perform the image stacking, we mask all the unrelated sources (in particular galaxies at low impact parameters having inconsistent photometric redshifts) present in the field so that the rms of the background counts in the final stacked image can be estimated accurately. We used the iterative $\sigma$-clipping method to estimate the mean (nearly zero) and the standard deviation ($\sigma_{bkg}$) of the background for every image and masked all the pixel having values more the 3$\sigma_{bkg}$ above the mean background and replace them with the mean background values. Once the masking of all the GOTOQ fields is performed, we visually inspected them to ensure that we have masked all the other sources except for the GOTOQs and the associated photometric extensions in the case of objects in the GOTOQ-Ext sub-sample. We then performed a simple median stacking of images separately for each of the above-mentioned three photometric bands. 

To estimate the average fluxes of the foreground galaxies in different bands we need to subtract the contributions of the quasar and its host galaxy from the stacked image. This we do by using the stacked images of an appropriately chosen control sample of quasars. We used quasars in the SDSS sample to construct the control sample. For every quasar in our GOTOQ sample, we randomly identified 5 quasars satisfying the following criteria: (i) Similar emission redshifts, i.e., $|\Delta z| \leqslant 0.005$; (ii) Similar SDSS r-band apparent magnitude, i.e., $|\Delta m_r| \leqslant 0.25$ mag and (iii) No \MgII\  absorption (having $\rm{W_{2796}} \geqslant 0.5$\AA) detected in the quasar spectra with at least $3\sigma$ significance. Except for five cases, where we could not find five quasars satisfying all three conditions, we slightly relaxed the first two conditions to find a total of five quasars with similar redshifts and apparent magnitudes and without any \MgII\ absorption present along the line of sight. Therefore, corresponding to 198 GOTOQs in our sample, there are a total of 990 quasars in our control sample. 

In Figure \ref{fig:compare_samples}, we compare the redshift  and the apparent magnitude distributions of quasars in our GOTOQ sample and in the control sample. The KS-test between the redshift and $m_r$ distributions of the GOTOQs and the control sample yields $p$ values of 1.00 and 0.92
respectively. This reconfirms that both samples are drawn from the same parent population as far as these two properties are concerned.
The $g,\, r,\, \text{and}\, z$ band images of these 990 quasars were also obtained from the DESI legacy survey. As in the case of GOTOQs, we first masked all the other sources present in the image and  performed the stacking to obtain the median stacked images of the control sample in all three different bands.

\begin{table*}
    \centering
   \caption{Estimated average galaxy properties based on the photometric and spectroscopic stacking. {The values in the parenthesis correspond to the $1\sigma$ uncertainty. Line luminosities are measured in the units of $10^{40}\, \rm{ergs\, s^{-1}}$}.}
   \begin{tabular}{ccccccccr}
    \hline
         Redshift & Sample & $m_g$ & $m_r$ & $m_z$ & $L_{[O\, \textsc{ii}]}$ & $L_{H\beta}$ & $L_{[O\, \textsc{iii}]}$ & $L_{[O\, \textsc{iii}]}$\\
                  &        &       &       &       & $\lambda 3728$ & $\lambda 4862$ & $\lambda 4960$ & $\lambda 5008$ \\
         \hline
         0.35 - 0.625 & All & $22.97^{+0.56}_{-0.89}$ & $21.80^{+0.23}_{-0.26}$ & $22.16^{+0.48}_{-0.65}$ & 9.98(0.32) & 3.48(0.25) & 2.57(0.27) & 6.93(0.25)\\
           (low-$z$)  & GOTOQ-Ext & $22.63^{+0.85}_{-0.95}$ & $21.43^{+0.40}_{-0.68}$ & $21.04^{+0.26}_{-0.37}$ & 11.2(0.5) & 4.0(0.4) & 2.9(0.4) & 11.0(0.5)\\
                      & GOTOQ-Noext  & $22.90^{+0.65}_{-0.97}$ & $22.32^{+0.31}_{-0.82}$ & $22.11^{+0.39}_{-0.71}$ & 9.7(0.4) & 2.96(0.31) & 2.75(0.31) & 5.70(0.31)\\
        \\         
        0.625 - 0.748 & All & $23.39^{+0.59}_{-0.85}$ & $22.37^{+0.35}_{-0.58}$ & $22.89^{+0.44}_{-0.93}$ & 18.9(0.7) & 3.8(0.6) & 4.2(0.9) & 10.9(0.8)\\
        (mid-$z$)  & GOTOQ-Ext & $22.38^{+0.73}_{-1.03}$ & $22.27^{+0.58}_{-0.95}$ & $22.81^{+0.53}_{-0.83}$ & 19.8(0.8) & 4.3(0.9) & 5.7(1.2) & 11.9(1.2)\\
        & GOTOQ-Noext & $22.63^{+0.97}_{-0.60}$ & $22.34^{+0.62}_{-0.68}$ & $22.49^{+0.70}_{-0.70}$ & 19.7(1.0) & $\leqslant$ 2.7 & 4.3(1.3) & 10.6(1.4) \\
        \\                  
        0.748 - 1.06  & All & $24.00^{^+0.59}_{-1.10}$ & $22.78^{+0.51}_{-0.70}$ & $23.07^{+0.52}_{-1.12}$ & 25.5(0.8) & 9.5(1.3) & $\leqslant$ 7.2 & 22.4(2.4)\\
        (high-$z$) & \\
        \hline
    \end{tabular}
 
    \label{tab:galaxy_props}
\end{table*}

\subsubsection{Host galaxy luminosity and impact parameter}

The GOTOQs in our sample spread over a wide redshift range. For the lowest redshift GOTOQ ($z_{abs}$ = 0.3662), one pixel in the image corresponds to a physical scale of 1.37 kpc while that for the highest redshift ($z_{abs}$ = 1.0582) GOTOQ is 2.19 kpc. To minimize the effect of angular to physical scale dependence on redshift,  we sub-divide our sample into three redshift bins (with an equal number of objects for the GOTOQ-Ext sample in each bin) and performed the image stacking of GOTOQs and the corresponding control sample QSOs.

\begin{figure*}
    \centering
    \includegraphics[viewport=8 8 510 400, width=\textwidth,clip=true]{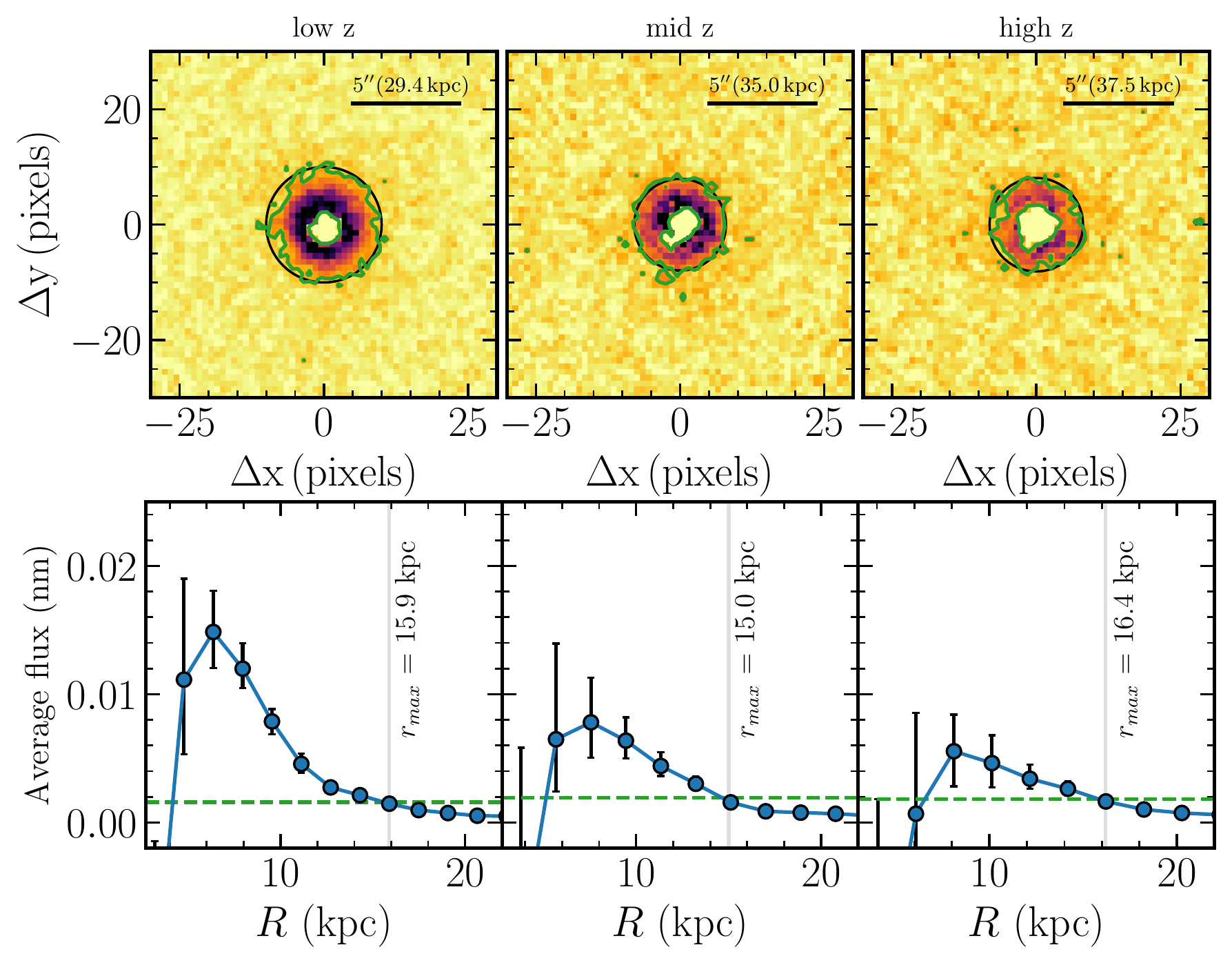}
    \caption{\emph{Top panel:} The  $r$-band stacked images of the GOTOQs in three redshift bins after subtracting the contributions from the QSO and its host galaxy using the control sample of quasars. 
    The 3$\sigma_{bgr}$ contours are shown in green.
    From the left to the right, the three panels correspond to the low-$z$, mid-$z$, and high-$z$ bins (see Table~\ref{tab:galaxy_props}) respectively. \emph{Bottom panel:} The radially averaged flux profiles (in nanomaggy) of the images shown in the top panel as a function of radial distance. The shown errors are obtained using bootstrapping. The green dashed horizontal line corresponds to the 3$\sigma_{bgr}$. The solid gray vertical line shows the radial distance where the flux falls below the 3$\sigma_{bgr}$. The black circles in the top panel indicate the radius (in the angular scale) corresponding to the $r_{max}$.}
    \label{fig:allz_rband}
\end{figure*}

\begin{figure*}
    \centering
    \includegraphics[viewport=7 8 510 395,width=\textwidth,clip=true]{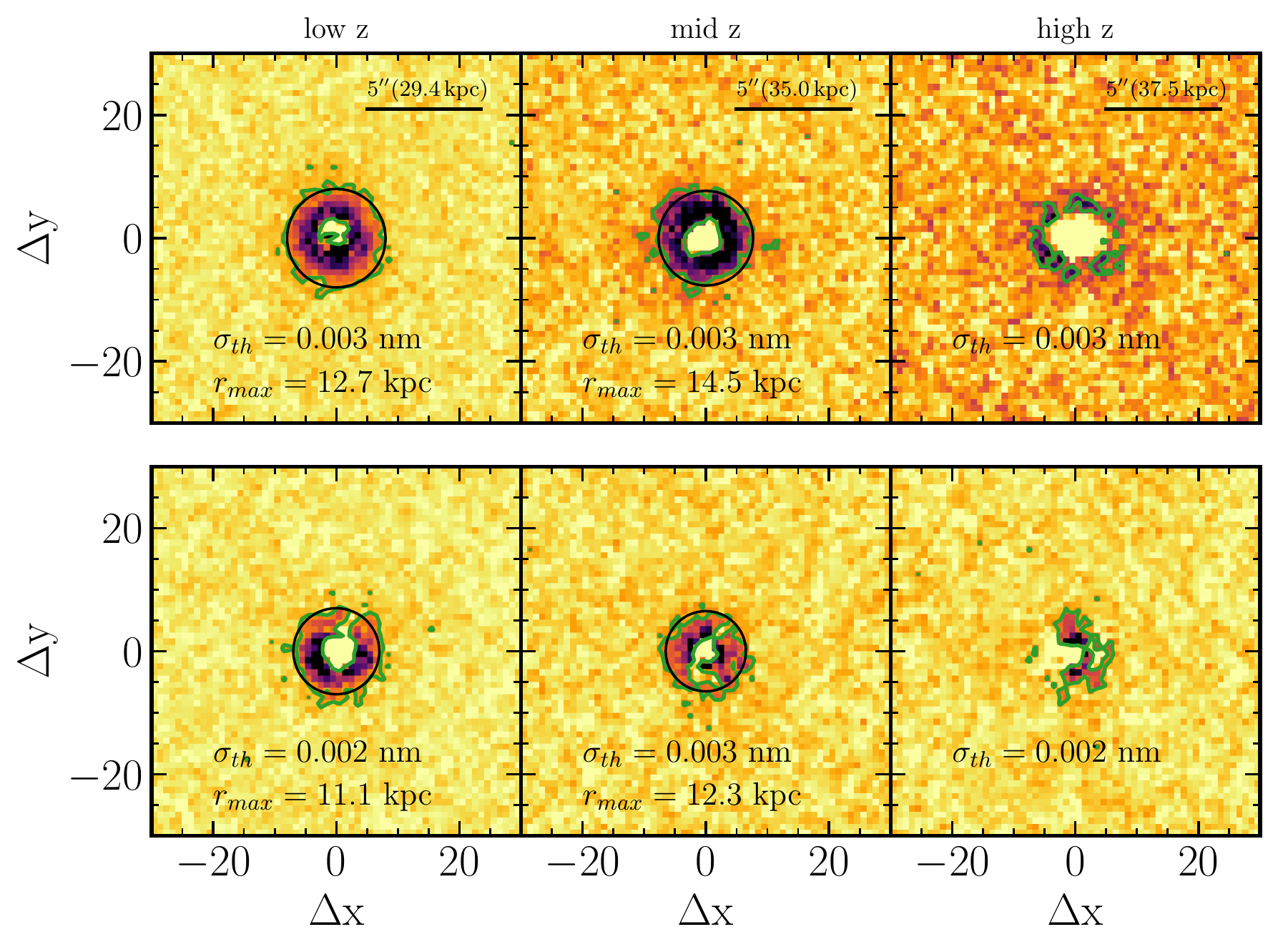}
    \caption{Residual r-band fluxes for the three redshift bins. The top and bottom panels correspond to the ``GOTOQ-Ext" and ``GOTOQ-Noext" sub-samples. The black circle corresponds to the radius of $r_{max}$ beyond which the average residual flux falls below the $\sigma_{th}$ (i.e. $3\times\sigma_{bgr}$). In each panel we provide $\sigma_{th}$ (in nanomaggy) and $r_{max}$.}
    \label{fig:extended_rad_profile}
\end{figure*}

In the top panel of Figure \ref{fig:allz_rband}, we show the r-band stacked images of the GOTOQs in three redshift bins after subtracting the corresponding images of the control sample. We clearly detect significant emissions from galaxies in all three redshift bins. The green contours in these panels show the 3 $\sigma_{bkg}$ level of the background noise. In the bottom panel of Figure \ref{fig:allz_rband}, we show the radially averaged flux profiles as a function of radial distance. Green dotted lines give the $3\sigma_{bkg}$ value of the background. The error bars show the 16th to 84th percentiles range of the measured data points obtained with bootstrapping by removing and replacing 20\% of the quasars from the given sample with 100 realizations. The vertical gray line provides the radial distance up to which galaxy emission is detected above $3\sigma_{bkg}$ level (the same is shown as dark circles in upper panels on the angular scale). This radial profile, which we measure from the stacked spectrum, is influenced by the intrinsic light profile of individual galaxies, the impact parameter distribution of galaxies, and the point spread function (PSF) of the observations. Therefore, the maximum distance up to which we can detect galaxy light provides a conservative upper limit on the impact parameter.  The maximum impact parameters for the three redshift bins range from 15.0 kpc to 16.4 kpc.
These are consistent with our expectations based on the projected size of fibers used in the SDSS spectroscopy and what we measure in the case of individual objects in the "GOTOQ-Ext" sub-sample.

In all figures, we notice a reduction in the flux close to the center. Whether this reduction is due to the over-subtraction or lack of galaxies close to the impact parameter zero, we perform the following exercise. For each quasar in our GOTOQ sample, out of the available five quasars in the control sample, we randomly select four quasars. Next, we
constructed two different control sub-samples (designated as ‘A’ and ‘B’) with the first two quasars in the control sub-sample ‘A’ and the other two in the control sub-sample ‘B’. Then, we create the median stacks for these two control sub-samples and subtract one from the other. The obtained residual is more or less consistent with zero and without any significant circular regions with negative counts in the center.  Thus, the ‘hole’ present in the stacks of GOTOQs is most likely to be related to the lack of galaxies close to zero impact parameters. Note presence of such GOTOQs would have considerably reddened the quasar, and probably the color selection used in SDSS would have missed such highly reddened objects. We integrate the flux in all the pixels of the residual image where the detection is above the $3\sigma_{bgr}$ level (i.e., between the inner and outer green contours). These are summarized in Table~\ref{tab:galaxy_props}. We used this to obtain the average apparent magnitudes in the three bands used here and the absolute magnitude in the B-band ($M_B$, assuming the average galaxy SED discussed below). The median $M_B$ obtained in these three redshift bins corresponds to a $\sim0.3L_*$ galaxy \citep{Faber2007} at these redshifts.

Next, we perform the image stacking of objects in the ``GOTOQ-Ext" and ``GOTOQ-Noext" sub-samples in the same three redshift bins. The resulting residual images are shown in Figure~\ref{fig:extended_rad_profile}. In the top panel, we show the results for the ``GOTOQ-Ext" sample for the three redshift bins in the r-band. The absorbing galaxies are clearly detected in the low-$z$ and mid-$z$ bins. The detection is marginal for the high-$z$ bin.  The measured apparent magnitudes in different bands and $M_B$ are also summarized in Table~\ref{tab:galaxy_props}.
Next, we compare the mean magnitude ($m_r$ and $M_B$) obtained from the stacking (yellow star) with that from the individual measurements (red star) obtained for ``GOTOQ-Ext" (see Figure~\ref{fig:galaxy_correlations}). For the low-$z$ and mid-$z$ bins, the estimated r-band magnitudes from the stacked images are up to 0.5 mag brighter than the median magnitudes from the direct measurements for the ``GOTOQ-Ext" sub-sample. However, in the high-$z$ bin they are consistent with one another. The excess seen for the low-$z$ and mid-$z$ bins could be related to a possible over-subtraction of QSO+host galaxy contribution in individual images due to larger $\sigma_{bgr}$ compared to that in our stacked images. High spatial resolution images are needed to understand the origin of this difference. If Figure~\ref{fig:extended_rad_profile}, we also indicate the maximum impact parameters. While these values are consistent with direct measurements, they are slightly lower than what we find for the full sample (see Figure~\ref{fig:allz_rband}). This is mainly because of the increase in the $\sigma_{brg}$ in the case of the ``GOTOQ-Ext" sub-sample.

In the bottom panel of Figure~\ref{fig:extended_rad_profile}, we show the results for the ``GOTOQ-Noext" sub-samples. We detect the host galaxies at a highly significant level in both low-$z$ and mid-$z$ bins. As there are more objects in this sub-sample compared to ``GOTOQ-Ext" sub-sample the $\sigma_{bgr}$ are slightly better for the ``GOTOQ-Noext" sub-sample. Despite this, we measure the $r_{max}$ to be smaller than what we have found for ``GOTOQ-Ext" sub-sample. Also, as can be seen from Table~\ref{tab:galaxy_props} in the r-band (where we have the best SNR), the measured absolute magnitude in the case of ``GOTOQ-Noext" is less than that of ``GOTOQ-Ext" sub-sample. Based on this, we can infer that the non-detection of extended features in ``GOTOQ-Noext" is  due to the impact parameters being smaller and the galaxies being fainter in this sub-sample.

\subsection{Stacking of quasar spectra}
\label{sec:stack_spectra}
In this section, we focus on the average galaxy spectrum obtained after removing the quasar contribution with the help of spectral stacking. We use this and the stacked images to derive the average properties of the host galaxies of the GOTOQs. We adopt the following procedure for the spectral stacking exercise. We first scale every quasar spectrum by an appropriate factor so that the flux in the individual spectrum matches well with a reference r-band flux. Next, to avoid contamination from the \lya\ forest absorption, we consider the wavelength range of the quasar spectrum that satisfies,
\begin{align*}
    \lambda_{obs} \geqslant (1 + z_{qso}) \times 1216\text{\AA} \times \bigg{(}1 + \frac{10^4 \rm{\,km\,s^{-1}}}{c}\bigg{)} 
\end{align*}
with `$c$' being the speed of light. Doppler shift corresponding to 10000 \kms\ is applied to avoid the N~{\sc v} emission line and possible associated absorption lines. Finally, we de-redshift every quasar spectrum to the rest-frame wavelength of the \MgII\ absorption by keeping the total flux conserved and re-sample the spectra to a common wavelength axis with $\Delta \lambda = 1$\AA. 
As the rest wavelength range of individual quasar spectrum varies depending on the \MgII\ absorption redshift, not all quasars contribute to a given rest wavelength in the stacked spectrum. However, we note that almost all quasars in the GOTOQ sample contribute to the stacking over the rest wavelength range 2800-5000\AA.

\subsubsection{Rest equivalent widths of absorption lines and dust extinction}

First, we study the rest equivalent widths of different absorption lines present in the stacked spectrum. For this, we consider the geometric mean of the continuum normalized spectra.  We first exclude the wavelength range affected by the emission and absorption line features for the continuum normalization and use \textsc{pyqsofit} \citep{Guo2018} to fit the quasar continuum. The rest equivalent widths of different absorption lines detected in the stacked spectra for the full and different sub-samples are summarized in Table~\ref{tab:rew}.  We also compare these with the 
results obtained from the stacked spectra of strong \MgII\ absorbers (i.e., $W_{2796}>2.0\AA$) by \citet[][Y06]{york2006}, \CaII\ absorbers by \citet[][S15]{sardane2015} and H~{\sc i} 21-cm absorbers by \citet[][D17]{Dutta2017}. We chose these data sets as strong \MgII\ absorbers (based on $W_{2796}$vs.D anti-correlation), \CaII\ absorbers \citep[][who found stronger \OII\ emission in the composite spectra of \CaII\ absorbers compared to \MgII\ absorbers]{Wild2007} and \HI\ 21-cm absorbers \citep[][]{gupta2010,Dutta2017qgp}  tend to probe galaxies at smaller impact parameters as in GOTOQs.

The average rest equivalent widths of \MgII\ measured in our sample is much higher than that measured for the \CaII-selected and H~{\sc i} 21-cm absorbers. This is also the case for strong Fe~{\sc ii} transitions (at rest wavelengths 2344, 2374, 2383, 2586, and 2600\AA). While the composite spectrum of Y06 (of systems with $W_{2976}\ge 2$\AA) has similar \MgII\ equivalent widths, all the other strong transitions have equivalent widths lower than what we measure for GOTOQs. This could imply a large spread in  line-of-sight velocity (or the number of individual absorption components) in the case of GOTOQs compared to other populations of absorbers considered here. The rest equivalent widths of the strongest transitions of \MgII\ and \FeII, which are expected to be highly saturated, are higher for the ``GOTOQ-Ext" sub-sample compared to that of the ``GOTOQ-Noext" sub-sample.

Interestingly the Mn~{\sc ii}  equivalent widths we find for the GOTOQs are consistent with what has been measured for H~{\sc i} 21-cm absorbers and slightly larger (albeit within error) in the case of Ca~{\sc ii} selected absorbers. Unlike the singly ionized species of Mg and Fe, Mg~{\sc i} equivalent width is nearly identical among different sub-sample.  \citet{Dutta2017} have shown that, for the similar \MgII\ equivalent widths, the absorbers showing H~{\sc i} 21-cm absorption (i.e. DLAs with cold neutral gas) tend to show detectable Mn~{\sc ii} in the SDSS spectra. All this implies most of the GOTOQs will be DLAs \citep[As confirmed for $z<0.15$ by][]{Kulkarni2022} specifically having larger velocity spread in singly ionized species compared to systems selected based on other tracers.

\begin{table*}
    \centering
\caption{Rest equivalent widths of the spectroscopic transitions detected in the geometric mean stacked quasar spectrum of the GOTOQ. 
}
\begin{tabular}{lcccccc}
     \hline
    Transition & \multicolumn{6}{c}{Rest equivalent width (\AA)} \\
       & All    & GOTOQ-Ext  & GOTOQ-Noext & Y06 & S15 & D17\\
    \hline
    \FeII$\lambda$2249 & $\le$0.10 & $\le$0.18 & $\le$0.21 & 0.11$\pm$0.01 & 0.126$\pm$0.012 &0.15$\pm$0.04\\
    \FeII$\lambda$2260 & 0.10$\pm0.03$ & $\le$0.20 & $\le0.20$ &  0.11$\pm$0.01 &0.115$\pm$0.011& 0.13$\pm$0.04\\
    \FeII$\lambda$2344 & 1.69$\pm$0.08 & 1.76$\pm$0.13 & 1.63$\pm$0.10 & 1.20$\pm$0.01 & 1.140$\pm$0.009&1.06$\pm$0.04\\
    \FeII$\lambda$2374 & 1.16$\pm$0.08 & 1.27$\pm$0.11 & 1.06$\pm$0.06 & 0.70$\pm$0.01 & 0.751$\pm$0.009&0.85$\pm$0.07\\
    \FeII$\lambda$2382 & 2.03$\pm$0.03 & 2.03$\pm$0.10 & 2.03$\pm$0.11 & 1.60$\pm$0.01 & 1.398$\pm$0.009 & 1.69$\pm$0.07\\
    \FeII$\lambda$2586 & 1.49$\pm$0.07 & 1.57$\pm$0.06 & 1.40$\pm$0.06 & 1.14$\pm$0.01 & 1.115$\pm$0.008 & 1.23$\pm$0.06\\
    \FeII$\lambda$2600 & 1.84$\pm$0.06 & 1.94$\pm$0.09 & 1.71$\pm$0.07 & 1.67$\pm$0.01 & 1.472$\pm$0.008 & 1.68$\pm$0.06\\
    \MnII$\lambda$2576 & 0.26$\pm$0.04 & 0.36$\pm$0.10 & 0.22$\pm$0.06 & 0.14$\pm$0.01 & 0.226$\pm$0.010 &0.28$\pm$0.06\\
    \MnII$\lambda$2594 & 0.25$\pm$0.04 & 0.30$\pm$0.06 & 0.27$\pm$0.05 & 0.13$\pm$0.01 & 0.164$\pm$0.009&0.20$\pm$0.06\\
    \MnII$\lambda$2606 & 0.13$\pm$0.04 & $\le$0.21 & $\le$0.15 & 0.08$\pm$0.01         & 0.104$\pm$0.009&0.14$\pm$0.06\\
    \MgII$\lambda$2796 & 2.52$\pm$0.03 & 2.66$\pm$0.05 & 2.44$\pm$0.06 & 2.67$\pm$0.01 & 1.940$\pm$0.008 & 2.22$\pm$0.08\\
    \MgII$\lambda$2803 & 2.29$\pm$0.04 & 2.45$\pm$0.04 & 2.16$\pm$0.07 & 2.40$\pm$0.01 & 1.803$\pm$0.008&2.08$\pm$0.07\\
    \MgI$\lambda$2852 & 0.73$\pm$0.02 & 0.67$\pm$0.03 & 0.72$\pm$0.05 & 0.59$\pm$0.01  & 0.742$\pm$0.007&0.77$\pm$0.07\\
    \TiII$\lambda$3384 & 0.17$\pm$0.06 & 0.15$\pm$0.05 & $\le$ 0.17 & .....& 0.082$\pm$0.007& ....\\
    \CaII$\lambda$3934 & 0.34$\pm$0.06 & 0.29$\pm$0.05 & 0.37$\pm$0.08 &....& 0.703$\pm$0.006& ....\\
    \CaII$\lambda$3969 & 0.15$\pm$0.05 & $\le$0.15& 0.17$\pm$0.05 & .... & 0.418$\pm$0.006& ....\\
    \hline
    \end{tabular}

    \label{tab:rew}
\end{table*}

\begin{figure}
    \centering
    \includegraphics[width=0.51\textwidth]{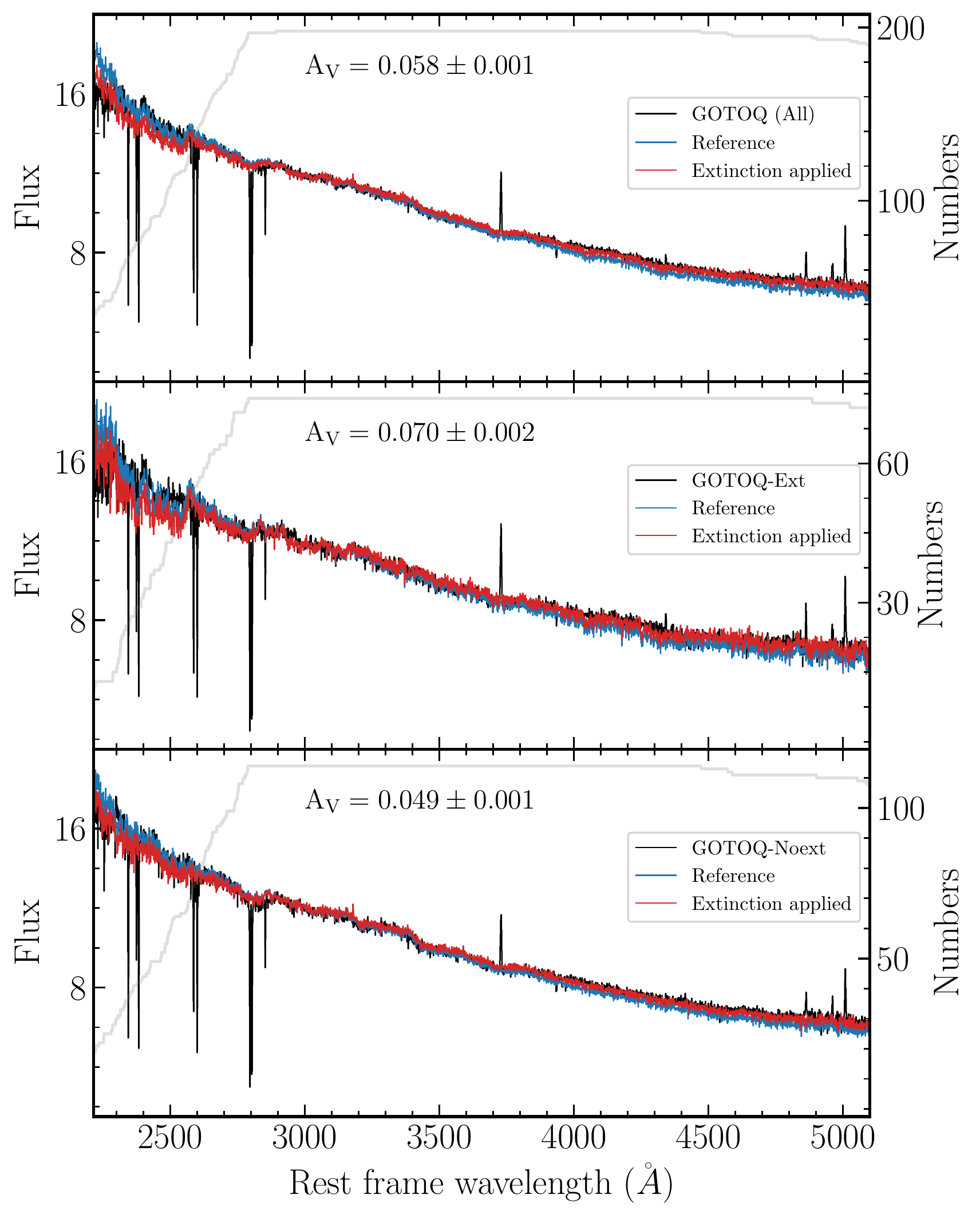}
    \caption{ Average reddening of the background quasars by all the GOTOQ in our full sample (top panel), ``GOTOQ-Ext" (middle panel) and ``GOTOQ-Noext" (bottom panel). The black spectrum in each panel is the geometric mean composite of the appropriate GOTOQ sample, the corresponding reference spectrum is shown in the blue and red spectrum corresponds to the best-fit reddened reference spectrum using the SMC extinction curve. The best-fitted $A_V$ values are indicated in each panel. {The gray line  gives to the number of quasars contributing to the stack at a given wavelength. }}
    \label{fig:dust_ext_noext}
\end{figure}
The \CaII\ absorption lines are clearly detected in our stacked spectrum. 
The measured equivalent width for the full GOTOQ sample is slightly higher than what is measured in the case of the strong \MgII\ absorber of Y06 but much weaker than what is found for \CaII-selected absorbers of S15. Among different sub-sample discussed by \citet{york2006} only the sub-sample of absorbers showing detectable Zn~{\sc ii} absorption have shown \CaII\ equivalent width close to what we find in our GOTOQ sample. 
\citet{Zhu2013ca2} have found a relationship between \CaII\ rest equivalent in the stacked spectrum with impact parameter using SDSS QSO-Galaxy pairs (at $z\sim0.1$) up to a projected distance of 200 kpc. For the impact parameter bin 3-10 kpc (with a median of 7 kpc) they found the rest equivalent width of \CaII$\lambda$3934 to be 0.435$\pm$0.068\AA. However, as in the case of \MgII\ discussed above a flattening in the W(\CaII) $vs.$ D relationship is noticed when individual measurements at D<15 kpc are considered \citep[][]{Straka_2015,Rubin2022}.  The fit by \citet{Rubin2022} predicts the W(\CaII) in the range 0.36-0.48\AA\ for D$<$14 kpc. The \CaII\ rest equivalent width we measure (i.e 0.34$\pm$0.06\AA) in our stacked spectrum is consistent with the above measurements at low redshifts. Another interesting thing we notice is that W(\CaII) obtained for the ``GOTOQ-Noext" sub-sample is slightly higher than that of the ``GOTOQ-Ext" sub-sample.

Next, we measure the average dust extinction using the geometric mean composite of the GOTOQ spectra as it is better suited to study the absorption properties. Using the composite spectrum of the control sample as a reference, we apply the SMC dust extinction law \citep{Gordon2003} and vary the V-band extinction coefficient, $A_V$ to match the continuum of the composite spectrum of the GOTOQ. Our method is similar to what is described in \citet{srianand2008, Guha2022}. The composite spectrum of the full GOTOQ sample, corresponding control sample and the best fit spectrum after applying extinction is shown in the top panel of  Figure \ref{fig:dust_ext_noext}. This resulted in the best fit value of $A_V = 0.058\pm0.001$ that corresponds to the color excess of $E(B-V) = 0.021\pm0.001$. This is consistent with what is found for the full \CaII-selected absorbers \citep[see][]{sardane2015} but less than what is found for the strong \CaII-selected absorbers and H~{\sc i} 21-cm absorbers \citep{Dutta2017}. The color excess we measure for GOTOQs is consistent with strong (2\AA $\leqslant W_{2796} \leqslant $ 6\AA) \MgII\ absorbers with high values ($\mathcal{R} \geqslant$ 0.5) of $W^{\rm{Fe\,\textsc{ii}}}_{2600} / W^{\rm{Mg\,\textsc{ii}}}_{2796}$  \citep{Joshi2018} as GOTOQ typically fall in this class of \MgII\ absorbers. In the middle and the bottom panels of Figure \ref{fig:dust_ext_noext}, we show the same for the ``GOTOQ-Ext" and the ``GOTOQ-Noext" sub-samples respectively. Compared to the ``GOTOQ-Noext" systems ($A_V = 0.049\pm0.001$ and E(B-V)$\sim$0.018), the ``GOTOQ-Ext" ($A_V = 0.070\pm0.002$; E(B-V)$\sim$ 0.025) systems are slightly more reddened. This trend is consistent with the known correlation between the \MgII\ equivalent width and E(B-V) \citep{Budzynski2011}.

\subsubsection{Average metallicity and ionization parameter of the nebular emission regions}

\begin{table*}
    \centering
  \caption{Measurement of gas-phase metallicity, the ionization parameters, and the stellar masses for the different GOTOQ samples based on emission line ratios and the SED fitting.}    \begin{tabular}{lccccr}
    \hline
    Sample & $Z$ & $q$ & Mass & SFR & Age\\
           & $12 + \log (O/H)$ & $\log(cm\, s^{-1})$ & $\log (M_\star / M_\odot)$ & $M_\odot / yr$ & Gyr\\
    \hline
    \multicolumn{4}{c}{low redshift bin}\\
    All & $8.32^{+0.09}_{-0.11}$ & $7.64^{+0.04}_{-0.09}$ & $9.76^{+0.10}_{-0.08}$ & $2.38^{+0.23}_{-0.26}$ & $3.97^{+0.86}_{-0.67}$\\
     GOTOQ-Ext & $8.31^{+0.10}_{-0.06}$ & $7.73^{+0.05}_{-0.08}$ & \\
     GOTOQ-Noext & $8.32^{+0.09}_{-0.08}$ & $7.62^{+0.04}_{-0.07}$ & \\
    
    \multicolumn{4}{c}{mid redshift bin}\\
    All & $8.32^{+0.02}_{-0.04}$ & $7.57^{+0.02}_{-0.05}$ & $9.46^{+0.14}_{-0.14}$ & $1.57^{+0.21}_{-0.24}$ & $3.07^{+1.14}_{-0.82}$\\
     GOTOQ-Ext & $8.32^{+0.06}_{-0.04}$ & $7.58^{+0.04}_{-0.03}$ & \\
     GOTOQ-Noext & -  & - & \\
    
    \multicolumn{4}{c}{high redshift bin}\\
    All & $8.31^{+0.09}_{-0.10}$ & $7.69^{+0.06}_{-0.10}$  & $9.81^{+0.10}_{-0.08}$ & $4.48^{+0.41}_{-0.40}$ & $2.59^{+0.54}_{-0.42}$\\
     GOTOQ-Ext & $8.31^{+0.29}_{-0.28}$ & $7.70^{+0.12}_{-0.14}$ & - \\
     GOTOQ-Noext & $8.31^{+0.06}_{-0.07}$ & $7.67^{+0.05}_{-0.08}$ & - \\
    
    \hline
    \end{tabular}
  
    \label{tab:zq}
\end{table*}
In this section, we use the average nebular emission line luminosities to derive the gas phase metallicity and ionization parameters. To estimate the median emission line luminosities, we first fitted continuum to individual spectrum using the package \textsc{pyqsofit} \citep{Guo2018}. Then, we subtracted each quasar continuum from their respective GOTOQ spectrum and then converted the fluxes to the line luminosities depending on the redshift of the foreground galaxies for the cosmological parameters assumed in this work. Then we performed a simple median stack of these systems. The median emission line luminosities for \OII, \OIII\, and H$\beta$ obtained from the stacked spectrum are listed in Table \ref{tab:galaxy_props}. The method used is very much similar to that of \citet{noterdaeme2010a,Menard2011} and \citet{Joshi2017}. As we do not incorporate the correction for fiber loss, the quoted average nebular line luminosities are lower limits. However, we proceed with the assumption that this correction factor is nearly the same for the nebular emission lines used. Therefore, fiber effects do not affect the line luminosity ratios that are crucial for deriving the parameters.

Based on the nebular emission line luminosities, we measure the ionization parameter ($q$) and the gas phase metallicity ($Z$) of the foreground galaxies using the \textsc{python} fork \citep{Mingozzi2020} of the \textsc{izi} \citep[Inferring metallicity and ionization parameter]{Blanc2015} and assuming the photoionization model of \citet{Levesque2010}. The measured values of $q$ and $Z$ are given in Table \ref{tab:zq}. It is evident that for the full as well as different sub-samples, the average derived metallicities are 12+log(O/H)$\sim$8.3 \citep[see also the Table 4 of][]{Joshi2017}, this is roughly a factor 2 less than the solar metallicity ($Z_\odot, \rm{(12 + \log (O/H))}$ = 8.69). The $q$ and $Z$ are similar for our GOTOQ-Ext and GOTOQ-Noext sub-samples.   We also did not find a clear evolution of these parameters with redshift over the small redshift range considered here.

Note that the average metallicity derived here is less than what is measured in the case of \usmg\ systems at similar redshifts by  \citet{Guha2022}. Even in few individual GOTOQs (with clear detections of all the nebular lines) where these measurements are made, the metallicities of GOTOQs are often sub-solar \citep[see figure 8 of][]{Guha2022}.
Based on the known stellar mass-metallicity relationship, this could imply the average stellar mass of the GOTOQs is lower than that of \usmg. 
If we use the Stellar mass- gas phase metallicity relationship found by \cite{Ma2016} we find the expected median stellar mass ($M_*$) to be in the range 10$^9$ to 3$\times10^{9}$ M$_\odot$. Below, we compare this with what we find through the SED fitting exercise.

The $q$ values measured for the different sub-samples of GOTOQs are in the range 7.5$\le log~q~(cm~s^{-1})\le$7.7.  These values are slightly higher than the mean $q$ value (i.e log~$q$ = 7.29$\pm$0.23) found for extra-galactic H~{\sc ii} regions in nearby spiral galaxies \citep{Blanc2015}. It is known in the literature that metallicity and $q$ are anti-correlated \citep{Kewley2019}. Thus slightly elevated $q$ values in our sample are expected as the average metallicity is a factor 2 less than solar.
\subsubsection{Median spectrum: Nature of galaxies}
\begin{figure*}
    \centering
    \includegraphics[viewport=10 5 615 690, width=0.9\textwidth,clip=true]{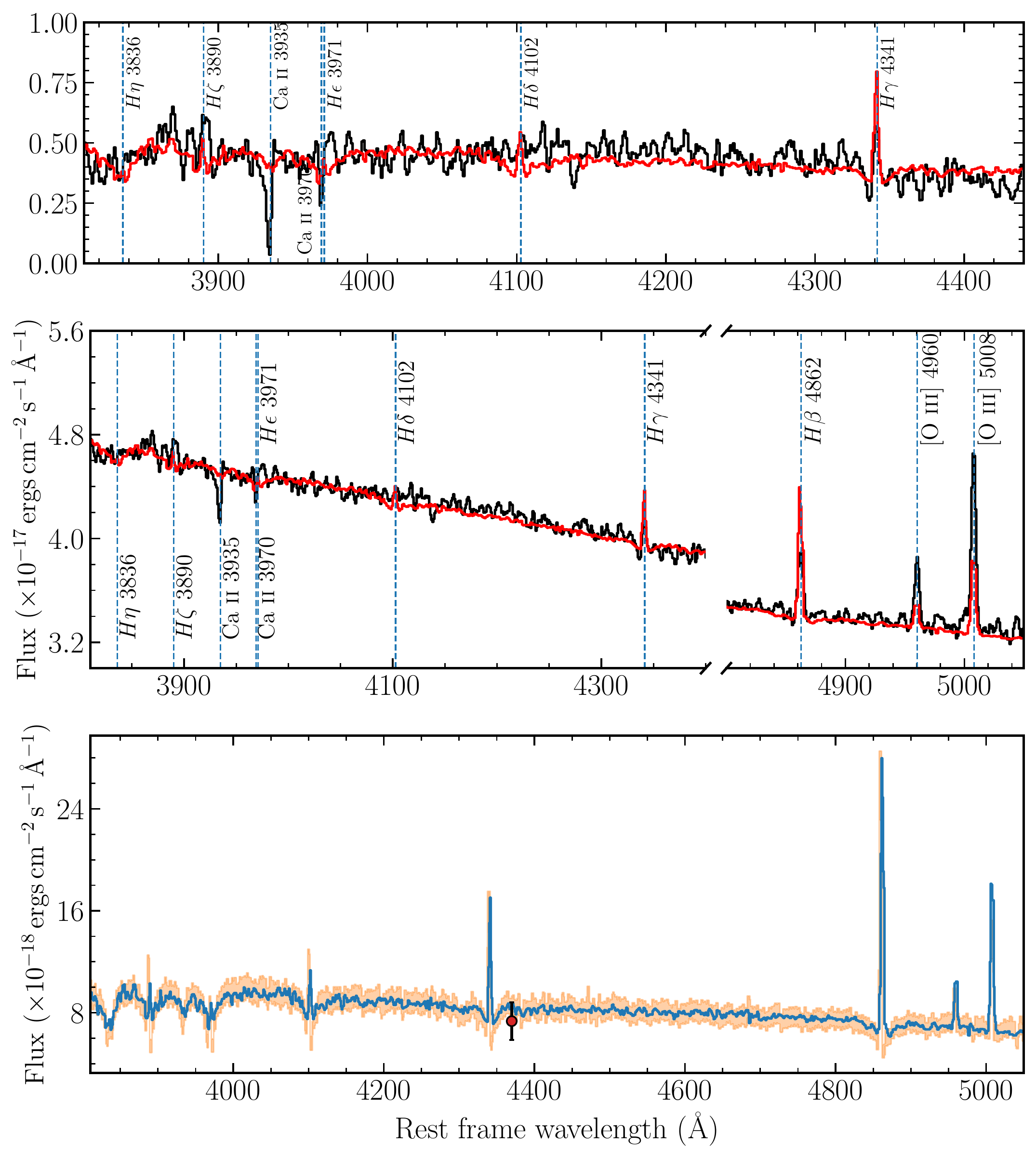}
    \caption{{\it Middle panel}: 
    The median stacked spectrum of the GOTOQs in the low-$z$ bin (black) and the best fit model(red) to this spectrum (power law + galaxy template) are shown. The blue dashed vertical lines show the expected locations of the prominent emission and absorption
    lines. \textit{Top panel:} The power-law subtracted stacked GOTOQ spectrum and the best fit model around the region containing \CaII\ absorption and Balmer lines from the  middle panel are shown. {\it Bottom panel}: The best-fitted galaxy template (blue) appropriately scaled to match the median flux (red point with error bar) seen in the stacked image of the low-$z$ sample.  The spectrum shown in orange is the SED fit that recovers all the nebular emissions, galactic absorption, and the stellar continuum features.}
    \label{fig:galspec}
\end{figure*} 
The median stacked spectrum obtained for the low-$z$ GOTOQ sample is shown in black in the middle panel of Figure \ref{fig:galspec}. In this panel, we also indicate the rest wavelengths of various emission and absorption features expected in the galaxy spectrum using blue vertical dashed lines. In addition to the nebular lines we have detected in individual cases (i.e., \OII, \OIII\, and H$\beta$), we detect H$\gamma$ and H$\delta$ in emission and higher Balmer lines in absorption in the composite spectrum. Note, because of fiber loss, we might have underestimated the emission line fluxes in individual cases and in the composite spectrum. However, as noted above, we assume that the line ratios are not severely affected by fiber loss.

The stacked spectrum shown has contributions from both the background quasars and the foreground galaxies. We assume the average contribution from the background quasars to be a power law and model the median stacked spectrum as a linear combination of the power law and a model galaxy template. For this, we use the model galaxy templates obtained from \citet{Fraix2021} who classified the  SDSS galaxies in 86 different spectroscopic classes using the k-means clustering algorithm. From these 86 spectroscopic classes, based on the least square minimization, we identify the best fit spectroscopic template. In this exercise, we have avoided the templates that might have contributions from AGN in addition to the stellar and nebular emissions.

In Figure \ref{fig:galspec}, the spectrum shown in red corresponds to the best-fitted galaxy template added to the power law contribution from quasars. The best-fitted galaxy template corresponds to the spectra class `B17' \citep{Fraix2021}, which are basically irregular star-forming galaxies. From the middle panel of Figure \ref{fig:galspec}, it is apparent that most of the Balmer absorption and emission lines present in the stacked GOTOQ spectrum are reproduced by this template. In the top panel, we show a zoomed-in power-law subtracted version of the fit from the middle panel to point out how well the galaxy template used  captures the Balmer absorption and emission lines. We notice that \CaII\ $\lambda\lambda$ 3935, 3970 absorption is much stronger than what has been seen in the template. We believe that considerable \CaII\ absorption originates from the CGM gas that produces the \MgII\ absorption along our line of sight. We repeat the same exercise for the three redshift bins and find that among the available star-forming galaxy templates, `B17' best fits the stacked GOTOQ spectrum for all the three redshift bins considered here. Although the `B17' template provides the best least square minimization, it does not very correctly produce the emission line ratios seen in the stacked spectrum. The model spectra of spectral class `A30' \citep{Fraix2021}, which corresponds to the composite spiral galaxies of Hubble type `S0', mimics the emission line ratios for all three redshift bins.

In the bottom panel of Figure \ref{fig:galspec},  We show the best-fitted galaxy template (in blue) scaled by the flux we measured in the stacked images for the low-z GOTOQs (given in Table~\ref{tab:galaxy_props}). We fit this best-fitting template (appropriately scaled to match the average flux measured in stacked images as shown in the bottom panel of Figure~\ref{fig:galspec}) using SED fitting techniques (discussed below) to derive parameters of the galaxies.

\subsubsection{Estaming Galaxy Parameters}
Using the three-band photometric measurements and our stacked spectra, we infer the average galaxy properties by fitting the spectral energy distribution (SED) of the galaxies in each of the redshift bins. For this purpose, we make use of the publicly available SED fitting code-named Bayesian Analysis of Galaxies for Physical Inference and Parameter EStimation \citep[\textsc{bagpipes}]{Carnall2018}. \textsc{bagpipes} assumes the stellar population synthesis models by \citet{Bruzual} and the implementations of the nebular emission lines from \citet{Byler}. Under the energy balance principle, we assume the \citet{1Calzetti1997} dust extinction law, such that dust-absorbed energy is re-radiated in the far infrared. All stellar populations have this effective absorption, while the youngest stars (defined as those with age < 10Myr) suffer an extra factor of attenuation ($\eta$, assumed to be 2.27) to account for dusty birth clouds. We also assume the delayed exponential star formation history. The obtained average stellar masses, the star formation rate, and the age of the GOTOQ in the three redshift bins are given in Table \ref{tab:zq}. The average stellar masses obtained for three redshift bins range from 9.46 to 9.81 while the SFR ranges from 1.57 to 4.48 $M_\odot\, yr^{-1}$. The stellar age  ranges from 2.59 Gyr to 3.97 Gyr. We note that the stellar mass derived by just fitting the SED to the three photometric points we have from the stacked images is also consistent with this value. Using the spectrum allows us to place additional constraints on age and SFR. For the 53 GOTOQs studied by \citet{Straka_2015}, the stellar masses ($\log\,  (M_\star / M_\odot)$) of the foreground galaxies range from 7.34 to 11.54 (with a median value of 9.05), and SFR ranges from 0.01 to 12 $M_\odot~yr^{-1}$ (with a median value of 0.54 $M_\odot~yr^{-1}$). Note that these SFR values should be considered as lower limits as they are based on H$\alpha$ luminosity that is affected by fiber size effects.

\begin{figure}
    \centering
    \includegraphics[viewport=8 8 360 335, width=0.475\textwidth,clip=true]{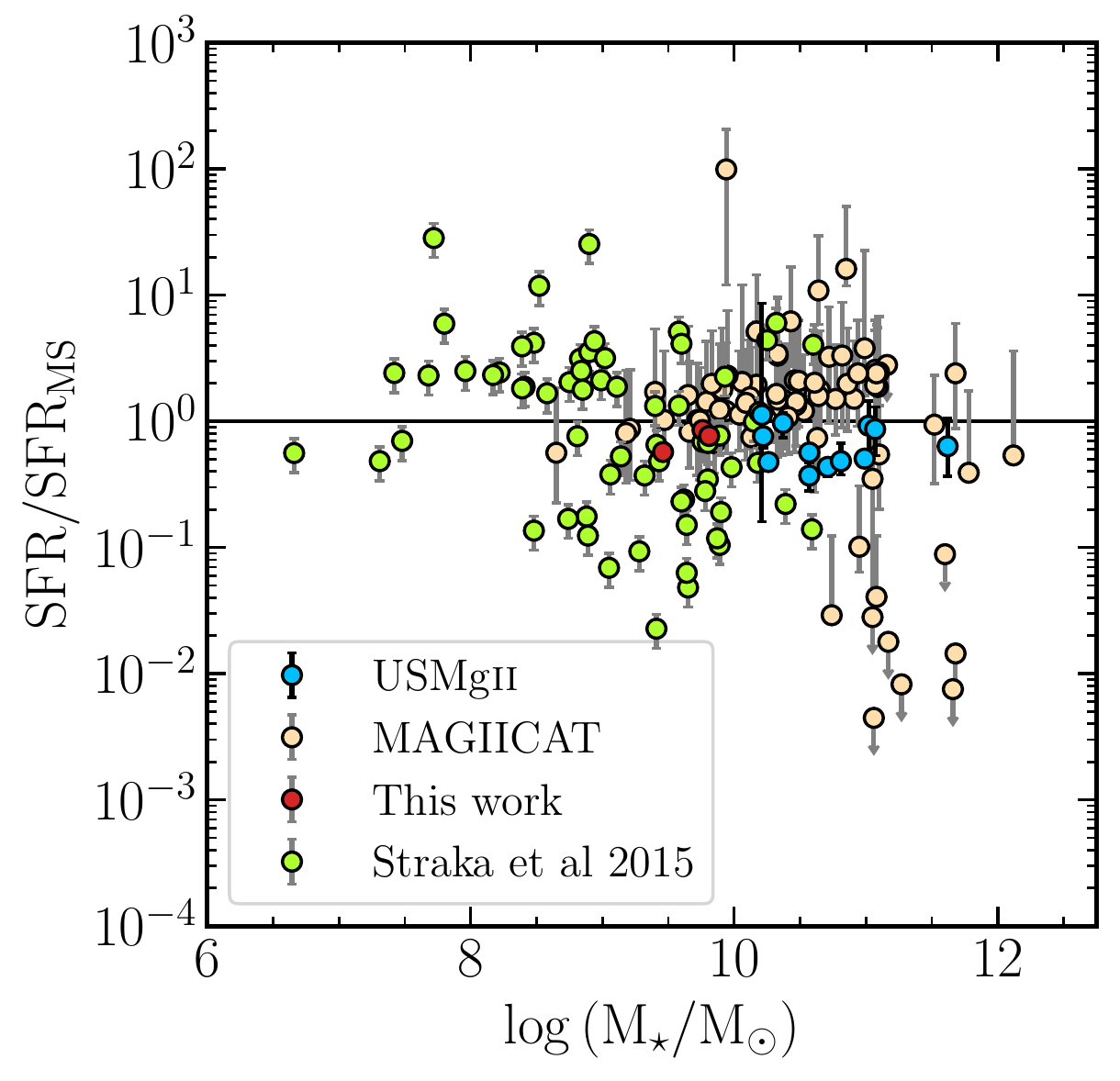}
    \caption{Comparison of the average properties of our GOTOQ host galaxy (red) from the three redshift bins with that of the population of normal \MgII\ absorbers (light orange, \textsc{magiicat}), GOTOQ \citep[light green]{Straka_2015}, and the \usmg\ (sky blue) systems. In the x-axis, we plot the stellar mass, and in the y-axis, we show the SFR scaled with respect to the main sequence SFR for the given stellar mass and the galaxy redshift. It is clear that the GOTOQ host galaxies have lower stellar mass but roughly follows the main sequence relations.}
    \label{fig:sfr_ms}
\end{figure}

In Figure \ref{fig:sfr_ms}, we plot the current star formation rate scaled with respect to the main sequence star-formation rate \citep{Speagle2014} for the same stellar mass and the galaxy redshift  vs. stellar mass for different samples of \MgII\ host galaxies. For the general population of the \MgII\ absorbers, we consider  objects in the well-known \textsc{magiicat} \citep{Nielsen_2013} sample of the \MgII\ absorption systems. \textsc{magiicat} survey is a compilation of a total of 182 \MgII\ absorption systems over the redshift ranges of $0.07 \leqslant z_{abs} \leqslant 1.12$ with the host galaxies identified up to a maximum impact parameter of about 200 kpc. To compare against the GOTOQ, however, we only consider the isolated \MgII\ host galaxies (defined as the only host galaxy lying within a maximum impact parameter of 100 kpc within a maximum velocity separation of 500 \kms within maximum impact parameters of 60 kpc. Among the 98 isolated \MgII\ host galaxies present within the impact parameter of 60 kpc in the \textsc{magiicat} survey, only 76 galaxies are identified as the photometric sources for which multiband reliable photometric analysis can be performed. For these 76 \MgII\ host galaxies, we have performed the SED fitting exercise using \textsc{bagpipes} to estimate the host galaxy properties like the stellar mass and the current star formation rate (SFR). The host galaxy properties of the \usmg\ host galaxies are taken from \citet{Guha2022} who compiled a sample of \usmg\ absorption systems accessible to the SALT ($\delta \leqslant 10^0$) within the redshift range of $0.4 \leqslant z_{abs} \leqslant 0.6$. We also show the results for GOTOQ from \citet{Straka_2015}. However, note that these points suffer from the aperture correction as SFR is estimated using H$\alpha$ luminosity.

In Figure~\ref{fig:sfr_ms},  the points for \usmg, \textsc{magiicat}, and the GOTOQ host galaxies are plotted in sky blue, light orange, and light green colors, respectively. {The average properties in the three redshift bins for the GOTOQ sample obtained from this work are shown in red.} The horizontal line corresponds to the main sequence star formation. Unlike the \usmg\ host galaxies, which are likely to be the post-starburst galaxies \citep{Guha2022}, the GOTOQ host galaxies are, on average, following the star-forming main sequence. It is also evident the average stellar mass (and SFR) found for the GOTOQs are on the lower side of what has been measured for the \textsc{magiicat} galaxies and much lower than what has been measured in the case of \usmg.

\section{Summary and Conclusions}
\label{sec:summary}

Frequency of occurrence and redshift evolution of \MgII\ absorption systems inferred from the spectra of high-$z$ quasars favor the existence of extended CGM around galaxies. Sustaining such a large reservoir of gas over cosmic time requires an equilibrium between outflowing gas from the galaxies and infalling gas into the galaxies. In particular, "down-the-barrel" spectroscopy of high-$z$ galaxies and spectroscopy of close pairs of background quasars and foreground galaxies are important to probe the nature of outflows at the interface between gaseous disk and CGM of galaxies. The host galaxies of GOTOQs and \usmg\ absorbers provide ideal targets to study quasar galaxy pairs at low impact parameters. Here, we study the nature of host galaxies of 198 GOTOQs in the sample of \citet{Joshi2017} using DESI Legacy Imaging Survey, SDSS spectra, image decomposition, and spectral and image stacking methods. Below, we summarize our findings.

\begin{itemize}
    \item [1.] We report the measurements of impact parameter ($D$) and absolute B-band magnitudes ($M_B$) for 74 systems, where decomposition of galaxy image from the quasar is possible in the DESI legacy survey images. The measured impact parameters are in the range 5.9$\le D[kpc]\le 16.9$. Our data alone does not show the well-known correlation between $W_{2796}$ and D.  However, our data combined with the literature data is well-fitted with a log-linear relationship (albeit with a large scatter in $W_{2796}$ for a given D). We measure $W_{2796}(D=0) = 3.44\pm 0.20$\AA\ and a exponential scale length of $21.6^{+2.41}_{-1.97}$ $kpc$ for the \MgII\ equivalent width profile.  We find neither of these parameters to show significant evolution with $z$ over $0.39\le z \le 1.05$. The value of $W_{2796}(D=0)$ obtained here is larger than what has been found by \citet{Kacprzak_2013}.
    This could just be due to their fit being weighted by a large number of measurements at large $D$, and an available handful of measurements at small $D$ not capturing the intrinsic scatter in the $W_{2796}$. 
    
    \citet{Joshi2017} have shown that the distribution of $W_{2796}$ in our sample (see their figure 7 and associated discussions in section 4.1) is statistically closer to what has been seen along ISM+halo sightlines of the Milky Way (MW) than that of absorption from the MW halo. \citet{Kacprzak_2013} found the same for absorbers with $D\le 6$ $kpc$ in their sample. \citet{Joshi2017} have also shown that the ratio of equivalent widths of Fe~{\sc ii}$\lambda$2600 and Mg~{\sc ii}$\lambda$2796 is systematically larger in our GOTOQs compared to the general population of \MgII\ absorbers. Our results support the model in which low impact parameter sightlines probe different populations of gas (disk+halo) compared to high impact parameter sightlines that mainly probe the halo gas. This picture is also endorsed by the fact that nearly all the GOTOQs are also either DLAs or sub-DLAs \citep[see][]{Kulkarni2022}. A smooth fit to the full sample using a log-linear fit may appear remarkable in this picture. But we wish to point out that this fit does have a large scatter, and it is important to look at properties other than $W_{2796}$ to probe distinction in the physical conditions of the gas at different radial distance scales from the galaxy. In this regard, measuring impact parameters in the remaining 124 GOTOQs in our sample using high spatial resolution observations will provide much tighter constraints on the $W_{2796}~vs.~D$ distribution at the low-D range.
   
    \item[2.]  The observed B-band absolute magnitudes are in the range $-22.34\le M_B\le-18.72$ mag. This corresponds to the rest frame B-band luminosity in the range 0.075$L_B^*$ to 2.07$L_B^*$ with $L_B^*$ being the characteristic  luminosity as defined in the Schechter function.
     We find a strong anti-correlation between $M_B$ and $D$ and a moderate correlation between $M_B$ and $W_{2796}$ in our sample, as has been the case for systems in the \textsc{magiicat} sample that typically probes larger impact parameters than our sample. These are consistent with brighter galaxies hosting larger CGM gas that produces large \MgII\ absorption line equivalent width (large velocity spread and the number of clouds), albeit with a large scatter.
     \item[3.] We performed an image stacking exercise to measure the average properties of host galaxies of all the 198 GOTOQs in our sample. To avoid the effect of evolving physical size for a fixed angular scale over redshift, we divide the sample into three redshift bins. Host galaxies are clearly detected in all three redshift bins and all three photometric bands used in the DESI legacy imaging survey. From the stacked images, we obtain the maximum impact parameter of our sample to be in the range $15\le D[kpc]\le16.4$ and the average B-band luminosity to be of the $\sim0.3~L_B^*$.  We also performed the image stacking exercise for the sub-samples where the host galaxies parameters are directly measurable (i.e. ``GOTOQ-Ext" sub-sample) and where the host  galaxies are not clearly visible (i.e. ``GOTOQ-Noext" sub-sample).   There is an indication for the impact parameters in the case of the ``GOTOQ-Noext" sub-sample being smaller than that of the ``GOTOQ-Ext" sub-sample. However, the average $M_B$ values inferred from the two samples are consistent with each other. Therefore, we conclude that the difficulty in detecting host galaxies in the case of the "GOTOQ-Nonext" sub-sample is probably related to the impact parameters being less in these cases. The lack of any difference in the distribution of $W_{2796}$ between the two sub-samples is also consistent with the flattening of $W_{2796}~vs. ~D $ distribution with a large scatter at small D.
     \item[4.] We performed different spectral stacking methods (i.e., median and geometric mean combined spectra) on the SDSS spectra to probe the average equivalent widths of CGM absorption lines, E(B$-$V), and spectral signatures (nebular emission and stellar absorption) from the host galaxies.  The absorption lines of strong transitions of \MgII\ and \FeII\ in the GOTOQ sample are found to be stronger than what has been seen in strong \MgII\ absorbers, Ca~{\sc ii} and H~{\sc i} 21-cm absorption selected systems. However, rest equivalent widths of Mg~{\sc i} and weaker transitions of \FeII (such as at 2249\AA\ and 2260\AA) lines are similar to what has been found in the above-mentioned absorbers. 
     We also find an average E(B-V) = $0.021\pm0.001$ for our sample. This is consistent with what is found for the full \CaII-selected absorbers \citep{sardane2015} but less than what is found for the strong \CaII-selected absorbers and \HI\ 21-cm absorbers\citep{Dutta2017}. 
     \item[5.] In the median stack of continuum-subtracted spectra, we detect nebular emission lines. We use the \textsc{ISI} code of \citet{Blanc2015} to estimate the average metallicity (Z) and ionization parameter ($q$) in three redshift bins. The inferred average metallicity (12+log($O/H$)$\sim$8.3) is less than the solar metallicity and corresponds to a stellar mass of $1-3\times 10^9$ M$_\odot$ if we use the known stellar mass-metallicity relationship at these redshifts. The $q$ values measured for the different sub-samples  are slightly higher than the mean $q$ value (i.e. log~$q\sim 7.29\pm0.23$) found for extra-galactic H~{\sc ii} regions in nearby spiral galaxies \citep{Blanc2015}. It is known in the literature that the metallicity and $q$ are anti-correlated \citep{Kewley2019}. Thus slightly elevated $q$ values in our sample are expected as the average metallicity is a factor 2 less than solar metallicity.
     \item[6.] We show that the median stacked spectrum can be well fitted by a QSO contribution approximated by a power-law plus the galaxy template `B17' of \citet{Fraix2021}. This template roughly follows the transition from emission to absorption in the case of Balmer lines, as observed in our stacked spectrum. This suggests on average GOTOQ host galaxies are irregular star-forming galaxies. We use the best-fitted galaxy template spectrum and  magnitudes obtained from the image stacking in three redshift bins  to derive the average star formation rate, stellar mass, and age using SED fitting. The average stellar mass inferred for the GOTOQs is typically less than what has been measured in the case of \MgII\ absorbers in the \textsc{magiicat} and \usmg\ samples. 
     We find that the average SFR inferred is consistent with what is expected for the main sequence based on the inferred average stellar mass. 
\end{itemize}

High spatial resolution imaging of our full sample (either with space observatories or using adaptive optics from ground-based facilities) is important to directly measure the impact parameter and orientation of galaxies for all the GOTOQs in our sample. Such observations will allow us to study the $W_{2796}~vs.~D$ correlations at $D\le 10$ kpc values and quantify the effect of galaxy orientation on the absorption profile. High spatial resolution spectroscopic observations will allow us to probe the absorbing gas toward both quasars and galaxies (i.e., down-the-barrel). Such observations, which will become possible with next-generation optical telescopes, will provide important insights into the gas distribution  and different feedback processes operating at $\le$10 $kpc$ to the high-$z$ star-forming galaxies. 

\section{Acknowledgement}
We thank the anonymous referee, Dr. Jens-Kristian Krogager, Dr. Rajeshwari Dutta, for their comments on the manuscript. This project makes use of the following softwares : NumPy \citep{numpy2020}, SciPy \citep{scipy2020}, Matplotlib \citep{matplotlib2007}, and AstroPy \citep{astropy:2013, astropy:2018}.
This paper makes use of SDSS observational data. Funding for the Sloan Digital Sky Survey IV has been provided by the Alfred P. Sloan Foundation, the U.S. Department of Energy Office of Science, and the Participating Institutions. SDSS-IV acknowledges support and resources from the Center for High-Performance Computing  at the University of Utah. The SDSS website is www.sdss.org. SDSS-IV is managed by the Astrophysical Research Consortium for the Participating Institutions of the SDSS Collaboration, including  the Brazilian Participation Group, the Carnegie Institution for Science, Carnegie Mellon University, Center for Astrophysics | Harvard \& Smithsonian, the Chilean Participation 
Group, the French Participation Group, Instituto de Astrof\'isica de 
Canarias, The Johns Hopkins University, Kavli Institute for the 
Physics and Mathematics of the Universe (IPMU) / University of Tokyo, the Korean Participation Group, Lawrence Berkeley National Laboratory, Leibniz Institut f\"ur Astrophysik Potsdam (AIP),  Max-Planck-Institut 
f\"ur Astronomie (MPIA Heidelberg), Max-Planck-Institut f\"ur Astrophysik (MPA Garching), Max-Planck-Institut f\"ur Extraterrestrische Physik (MPE), National Astronomical Observatories of China, New Mexico State University, 
New York University, University of Notre Dame, Observat\'ario Nacional / MCTI, The Ohio State University, Pennsylvania State University, Shanghai 
Astronomical Observatory, United Kingdom Participation Group, 
Universidad Nacional Aut\'onoma de M\'exico, University of Arizona, University of Colorado Boulder, University of Oxford, University of Portsmouth, University of Utah, University of Virginia, University of Washington, University of Wisconsin, Vanderbilt University, and Yale University.

The DESI Legacy Imaging Surveys consist of three individual and complementary projects: the Dark Energy Camera Legacy Survey (DECaLS), the Beijing-Arizona Sky Survey (BASS), and the Mayall z-band Legacy Survey (MzLS). DECaLS, BASS, and MzLS together include data obtained, respectively, at the Blanco telescope, Cerro Tololo Inter-American Observatory, NSF’s NOIRLab; the Bok telescope, Steward Observatory, University of Arizona; and the Mayall telescope, Kitt Peak National Observatory, NOIRLab. NOIRLab is operated by the Association of Universities for Research in Astronomy (AURA) under a cooperative agreement with the National Science Foundation. Pipeline processing and analyses of the data were supported by NOIRLab and the Lawrence Berkeley National Laboratory (LBNL). Legacy Surveys also uses data products from the Near-Earth Object Wide-field Infrared Survey Explorer (NEOWISE), a project of the Jet Propulsion Laboratory/California Institute of Technology, funded by the National Aeronautics and Space Administration. Legacy Surveys was supported by: the Director, Office of Science, Office of High Energy Physics of the U.S. Department of Energy; the National Energy Research Scientific Computing Center, a DOE Office of Science User Facility; the U.S. National Science Foundation, Division of Astronomical Sciences; the National Astronomical Observatories of China, the Chinese Academy of Sciences and the Chinese National Natural Science Foundation. LBNL is managed by the Regents of the University of California under contract to the U.S. Department of Energy. The complete acknowledgments can be found at https://www.legacysurvey.org/acknowledgment/. The Photometric Redshifts for the Legacy Surveys (PRLS) catalog used in this paper was produced thanks to funding from the U.S. Department of Energy Office of Science, Office of High Energy Physics via grant DE-SC0007914.
\label{sec:acknowledgement}

\section{Data Availability}
{All the spectroscopic and the photometric data used in this work are publicly available from the SDSS\footnote{https://www.sdss.org/} and the DESI Legacy Imaging Surveys respectively\footnote{https://www.legacysurvey.org/}.}



\bibliographystyle{mnras}
\bibliography{example} 

\begin{thebibliography}{}
\makeatletter
\relax
\def\mn@urlcharsother{\let\do\@makeother \do\$\do\&\do\#\do\^\do\_\do\%\do\~}
\def\mn@doi{\begingroup\mn@urlcharsother \@ifnextchar [ {\mn@doi@}
  {\mn@doi@[]}}
\def\mn@doi@[#1]#2{\def\@tempa{#1}\ifx\@tempa\@empty \href
  {http://dx.doi.org/#2} {doi:#2}\else \href {http://dx.doi.org/#2} {#1}\fi
  \endgroup}
\def\mn@eprint#1#2{\mn@eprint@#1:#2::\@nil}
\def\mn@eprint@arXiv#1{\href {http://arxiv.org/abs/#1} {{\tt arXiv:#1}}}
\def\mn@eprint@dblp#1{\href {http://dblp.uni-trier.de/rec/bibtex/#1.xml}
  {dblp:#1}}
\def\mn@eprint@#1:#2:#3:#4\@nil{\def\@tempa {#1}\def\@tempb {#2}\def\@tempc
  {#3}\ifx \@tempc \@empty \let \@tempc \@tempb \let \@tempb \@tempa \fi \ifx
  \@tempb \@empty \def\@tempb {arXiv}\fi \@ifundefined
  {mn@eprint@\@tempb}{\@tempb:\@tempc}{\expandafter \expandafter \csname
  mn@eprint@\@tempb\endcsname \expandafter{\@tempc}}}

\bibitem[\protect\citeauthoryear{{Astropy Collaboration} et~al.,}{{Astropy
  Collaboration} et~al.}{2013}]{astropy:2013}
{Astropy Collaboration} et~al., 2013, \mn@doi [\aap]
  {10.1051/0004-6361/201322068}, \href
  {http://adsabs.harvard.edu/abs/2013A%26A...558A..33A} {558, A33}

\bibitem[\protect\citeauthoryear{{Astropy Collaboration} et~al.,}{{Astropy
  Collaboration} et~al.}{2018}]{astropy:2018}
{Astropy Collaboration} et~al., 2018, \mn@doi [\aj] {10.3847/1538-3881/aabc4f},
  \href {https://ui.adsabs.harvard.edu/abs/2018AJ....156..123A} {156, 123}

\bibitem[\protect\citeauthoryear{{Blanc}, {Kewley}, {Vogt}  \&
  {Dopita}}{{Blanc} et~al.}{2015}]{Blanc2015}
{Blanc} G.~A.,  {Kewley} L.,  {Vogt} F. P.~A.,   {Dopita} M.~A.,  2015, \mn@doi
  [\apj] {10.1088/0004-637X/798/2/99}, \href
  {https://ui.adsabs.harvard.edu/abs/2015ApJ...798...99B} {798, 99}

\bibitem[\protect\citeauthoryear{Bond, Churchill, Charlton  \& Vogt}{Bond
  et~al.}{2001}]{bond2001high}
Bond N.~A.,  Churchill C.~W.,  Charlton J.~C.,   Vogt S.~S.,  2001, The
  Astrophysical Journal, 562, 641

\bibitem[\protect\citeauthoryear{{Bordoloi}, {Lilly}, {Kacprzak}  \&
  {Churchill}}{{Bordoloi} et~al.}{2014}]{Bordoloi2014}
{Bordoloi} R.,  {Lilly} S.~J.,  {Kacprzak} G.~G.,   {Churchill} C.~W.,  2014,
  \mn@doi [The Astrophysical Journal] {10.1088/0004-637X/784/2/108}, \href
  {https://ui.adsabs.harvard.edu/abs/2014ApJ...784..108B} {784, 108}

\bibitem[\protect\citeauthoryear{Bruzual \& Charlot}{Bruzual \&
  Charlot}{2003}]{Bruzual}
Bruzual G.,  Charlot S.,  2003, \mn@doi [Monthly Notices of the Royal
  Astronomical Society] {10.1046/j.1365-8711.2003.06897.x}, 344, 1000

\bibitem[\protect\citeauthoryear{{Budzynski} \& {Hewett}}{{Budzynski} \&
  {Hewett}}{2011}]{Budzynski2011}
{Budzynski} J.~M.,  {Hewett} P.~C.,  2011, \mn@doi [\mnras]
  {10.1111/j.1365-2966.2011.19158.x}, \href
  {https://ui.adsabs.harvard.edu/abs/2011MNRAS.416.1871B} {416, 1871}

\bibitem[\protect\citeauthoryear{{Byler}, {Dalcanton}, {Conroy}  \&
  {Johnson}}{{Byler} et~al.}{2017}]{Byler}
{Byler} N.,  {Dalcanton} J.~J.,  {Conroy} C.,   {Johnson} B.~D.,  2017, \mn@doi
  [\apj] {10.3847/1538-4357/aa6c66}, \href
  {https://ui.adsabs.harvard.edu/abs/2017ApJ...840...44B} {840, 44}

\bibitem[\protect\citeauthoryear{{Calzetti}}{{Calzetti}}{1997}]{1Calzetti1997}
{Calzetti} D.,  1997, \mn@doi [\aj] {10.1086/118242}, \href
  {https://ui.adsabs.harvard.edu/abs/1997AJ....113..162C} {113, 162}

\bibitem[\protect\citeauthoryear{Carnall, McLure, Dunlop  \& Davé}{Carnall
  et~al.}{2018}]{Carnall2018}
Carnall A.~C.,  McLure R.~J.,  Dunlop J.~S.,   Davé R.,  2018, \mn@doi
  [Monthly Notices of the Royal Astronomical Society] {10.1093/mnras/sty2169},
  480, 4379

\bibitem[\protect\citeauthoryear{Chen, Helsby, Gauthier, Shectman, Thompson  \&
  Tinker}{Chen et~al.}{2010}]{Chen_2010}
Chen H.-W.,  Helsby J.~E.,  Gauthier J.-R.,  Shectman S.~A.,  Thompson I.~B.,
  Tinker J.~L.,  2010, \mn@doi [The Astrophysical Journal]
  {10.1088/0004-637x/714/2/1521}, 714, 1521

\bibitem[\protect\citeauthoryear{{Dey}, {Torrey}, {Rubin}, {Zhu}  \&
  {Suresh}}{{Dey} et~al.}{2015}]{Dey2015}
{Dey} A.,  {Torrey} P.,  {Rubin} K. H.~R.,  {Zhu} G.~B.,   {Suresh} J.,  2015,
  \mn@doi [Monthly Notices of the Royal Astronomical Society]
  {10.1093/mnras/stv604}, \href
  {https://ui.adsabs.harvard.edu/abs/2015MNRAS.451.1806D} {451, 1806}

\bibitem[\protect\citeauthoryear{{Dey} et~al.,}{{Dey}
  et~al.}{2019}]{Dey2019DESI}
{Dey} A.,  et~al., 2019, \mn@doi [\aj] {10.3847/1538-3881/ab089d}, \href
  {https://ui.adsabs.harvard.edu/abs/2019AJ....157..168D} {157, 168}

\bibitem[\protect\citeauthoryear{{Dutta}, {Srianand}, {Gupta}, {Momjian},
  {Noterdaeme}, {Petitjean}  \& {Rahmani}}{{Dutta} et~al.}{2017a}]{Dutta2017}
{Dutta} R.,  {Srianand} R.,  {Gupta} N.,  {Momjian} E.,  {Noterdaeme} P.,
  {Petitjean} P.,   {Rahmani} H.,  2017a, \mn@doi [\mnras]
  {10.1093/mnras/stw2689}, \href
  {https://ui.adsabs.harvard.edu/abs/2017MNRAS.465..588D} {465, 588}

\bibitem[\protect\citeauthoryear{{Dutta}, {Srianand}, {Gupta}, {Momjian},
  {Noterdaeme}, {Petitjean}  \& {Rahmani}}{{Dutta}
  et~al.}{2017b}]{Dutta2017qgp}
{Dutta} R.,  {Srianand} R.,  {Gupta} N.,  {Momjian} E.,  {Noterdaeme} P.,
  {Petitjean} P.,   {Rahmani} H.,  2017b, \mn@doi [\mnras]
  {10.1093/mnras/stw2689}, \href
  {https://ui.adsabs.harvard.edu/abs/2017MNRAS.465..588D} {465, 588}

\bibitem[\protect\citeauthoryear{{Dutta} et~al.,}{{Dutta}
  et~al.}{2020}]{Dutta2020}
{Dutta} R.,  et~al., 2020, \mn@doi [\mnras] {10.1093/mnras/staa3147}, \href
  {https://ui.adsabs.harvard.edu/abs/2020MNRAS.499.5022D} {499, 5022}

\bibitem[\protect\citeauthoryear{{Dutta} et~al.,}{{Dutta}
  et~al.}{2021}]{Dutta2021}
{Dutta} R.,  et~al., 2021, \mn@doi [\mnras] {10.1093/mnras/stab2752}, \href
  {https://ui.adsabs.harvard.edu/abs/2021MNRAS.508.4573D} {508, 4573}

\bibitem[\protect\citeauthoryear{{Erb}}{{Erb}}{2008}]{Erb2008}
{Erb} D.~K.,  2008, \mn@doi [The Astrophysical Journal] {10.1086/524727}, \href
  {https://ui.adsabs.harvard.edu/abs/2008ApJ...674..151E} {674, 151}

\bibitem[\protect\citeauthoryear{{Faber} et~al.,}{{Faber}
  et~al.}{2007}]{Faber2007}
{Faber} S.~M.,  et~al., 2007, \mn@doi [\apj] {10.1086/519294}, \href
  {https://ui.adsabs.harvard.edu/abs/2007ApJ...665..265F} {665, 265}

\bibitem[\protect\citeauthoryear{{Fraix-Burnet}, {Bouveyron}  \&
  {Moultaka}}{{Fraix-Burnet} et~al.}{2021}]{Fraix2021}
{Fraix-Burnet} D.,  {Bouveyron} C.,   {Moultaka} J.,  2021, \mn@doi [\aap]
  {10.1051/0004-6361/202040046}, \href
  {https://ui.adsabs.harvard.edu/abs/2021A&A...649A..53F} {649, A53}

\bibitem[\protect\citeauthoryear{Gauthier}{Gauthier}{2013}]{Gauthier2013}
Gauthier J.-R.,  2013, \mn@doi [Monthly Notices of the Royal Astronomical
  Society] {10.1093/mnras/stt565}, \href {https://doi.org/10.1093/mnras/stt565}
  {432, 1444}

\bibitem[\protect\citeauthoryear{{Gordon}, {Clayton}, {Misselt}, {Landolt}  \&
  {Wolff}}{{Gordon} et~al.}{2003}]{Gordon2003}
{Gordon} K.~D.,  {Clayton} G.~C.,  {Misselt} K.~A.,  {Landolt} A.~U.,   {Wolff}
  M.~J.,  2003, \mn@doi [\apj] {10.1086/376774}, \href
  {https://ui.adsabs.harvard.edu/abs/2003ApJ...594..279G} {594, 279}

\bibitem[\protect\citeauthoryear{{Guha}, {Srianand}, {Dutta}, {Joshi},
  {Noterdaeme}  \& {Petitjean}}{{Guha} et~al.}{2022}]{Guha2022}
{Guha} L.~K.,  {Srianand} R.,  {Dutta} R.,  {Joshi} R.,  {Noterdaeme} P.,
  {Petitjean} P.,  2022, \mn@doi [\mnras] {10.1093/mnras/stac1106}, \href
  {https://ui.adsabs.harvard.edu/abs/2022MNRAS.513.3836G} {513, 3836}

\bibitem[\protect\citeauthoryear{{Guo}, {Shen}  \& {Wang}}{{Guo}
  et~al.}{2018}]{Guo2018}
{Guo} H.,  {Shen} Y.,   {Wang} S.,  2018, {PyQSOFit: Python code to fit the
  spectrum of quasars}, Astrophysics Source Code Library, record ascl:1809.008
  (\mn@eprint {ascl} {1809.008})

\bibitem[\protect\citeauthoryear{{Gupta}, {Srianand}, {Bowen}, {York}  \&
  {Wadadekar}}{{Gupta} et~al.}{2010}]{gupta2010}
{Gupta} N.,  {Srianand} R.,  {Bowen} D.~V.,  {York} D.~G.,   {Wadadekar} Y.,
  2010, \mn@doi [\mnras] {10.1111/j.1365-2966.2010.17198.x}, \href
  {http://adsabs.harvard.edu/abs/2010MNRAS.408..849G} {408, 849}

\bibitem[\protect\citeauthoryear{{Gupta}, {Srianand}, {Petitjean}, {Bergeron},
  {Noterdaeme}  \& {Muzahid}}{{Gupta} et~al.}{2012}]{gupta2012}
{Gupta} N.,  {Srianand} R.,  {Petitjean} P.,  {Bergeron} J.,  {Noterdaeme} P.,
   {Muzahid} S.,  2012, \mn@doi [\aap] {10.1051/0004-6361/201219159}, \href
  {http://adsabs.harvard.edu/abs/2012A%26A...544A..21G} {544, A21}

\bibitem[\protect\citeauthoryear{Harris et~al.,}{Harris
  et~al.}{2020}]{numpy2020}
Harris C.~R.,  et~al., 2020, \mn@doi [Nature] {10.1038/s41586-020-2649-2}, 585,
  357–362

\bibitem[\protect\citeauthoryear{Houck \& Bregman}{Houck \&
  Bregman}{1990}]{houck1990low}
Houck J.~C.,  Bregman J.~N.,  1990, The Astrophysical Journal, 352, 506

\bibitem[\protect\citeauthoryear{{Huang}, {Chen}, {Shectman}, {Johnson},
  {Zahedy}, {Helsby}, {Gauthier}  \& {Thompson}}{{Huang}
  et~al.}{2021}]{Huang2021}
{Huang} Y.-H.,  {Chen} H.-W.,  {Shectman} S.~A.,  {Johnson} S.~D.,  {Zahedy}
  F.~S.,  {Helsby} J.~E.,  {Gauthier} J.-R.,   {Thompson} I.~B.,  2021, \mn@doi
  [\mnras] {10.1093/mnras/stab360}, \href
  {https://ui.adsabs.harvard.edu/abs/2021MNRAS.502.4743H} {502, 4743}

\bibitem[\protect\citeauthoryear{Hunter}{Hunter}{2007}]{matplotlib2007}
Hunter J.~D.,  2007, \mn@doi [Computing in Science Engineering]
  {10.1109/MCSE.2007.55}, 9, 90

\bibitem[\protect\citeauthoryear{{Joshi}, {Srianand}, {Petitjean}  \&
  {Noterdaeme}}{{Joshi} et~al.}{2017}]{Joshi2017}
{Joshi} R.,  {Srianand} R.,  {Petitjean} P.,   {Noterdaeme} P.,  2017, \mn@doi
  [Monthly Notices of the Royal Astronomical Society] {10.1093/mnras/stx1499},
  \href {https://ui.adsabs.harvard.edu/abs/2017MNRAS.471.1910J} {471, 1910}

\bibitem[\protect\citeauthoryear{{Joshi}, {Srianand}, {Petitjean}  \&
  {Noterdaeme}}{{Joshi} et~al.}{2018}]{Joshi2018}
{Joshi} R.,  {Srianand} R.,  {Petitjean} P.,   {Noterdaeme} P.,  2018, \mn@doi
  [\mnras] {10.1093/mnras/sty121}, \href
  {https://ui.adsabs.harvard.edu/abs/2018MNRAS.476..210J} {476, 210}

\bibitem[\protect\citeauthoryear{{Kacprzak}}{{Kacprzak}}{2017}]{Kacprzak2017_gas_accretion}
{Kacprzak} G.~G.,  2017, in {Fox} A.,  {Dav{\'e}} R.,  eds,  Astrophysics and
  Space Science Library Vol. 430, Gas Accretion onto Galaxies. p.~145
  (\mn@eprint {arXiv} {1612.00451}), \mn@doi{10.1007/978-3-319-52512-9\_7}

\bibitem[\protect\citeauthoryear{Kacprzak, Cooke, Churchill, Ryan-Weber  \&
  Nielsen}{Kacprzak et~al.}{2013}]{Kacprzak_2013}
Kacprzak G.~G.,  Cooke J.,  Churchill C.~W.,  Ryan-Weber E.~V.,   Nielsen
  N.~M.,  2013, \mn@doi [The Astrophysical Journal]
  {10.1088/2041-8205/777/1/l11}, 777, L11

\bibitem[\protect\citeauthoryear{{Karademir} et~al.,}{{Karademir}
  et~al.}{2022}]{Karademir2022}
{Karademir} G.~S.,  et~al., 2022, \mn@doi [\mnras] {10.1093/mnras/stab3229},
  \href {https://ui.adsabs.harvard.edu/abs/2022MNRAS.509.5467K} {509, 5467}

\bibitem[\protect\citeauthoryear{{Kennicutt} \& {Evans}}{{Kennicutt} \&
  {Evans}}{2012}]{Kennicutt2012}
{Kennicutt} R.~C.,  {Evans} N.~J.,  2012, \mn@doi [\araa]
  {10.1146/annurev-astro-081811-125610}, \href
  {https://ui.adsabs.harvard.edu/abs/2012ARA&A..50..531K} {50, 531}

\bibitem[\protect\citeauthoryear{{Kewley}, {Nicholls}  \&
  {Sutherland}}{{Kewley} et~al.}{2019}]{Kewley2019}
{Kewley} L.~J.,  {Nicholls} D.~C.,   {Sutherland} R.~S.,  2019, \mn@doi [\araa]
  {10.1146/annurev-astro-081817-051832}, \href
  {https://ui.adsabs.harvard.edu/abs/2019ARA&A..57..511K} {57, 511}

\bibitem[\protect\citeauthoryear{{Kulkarni}, {Bowen}, {Straka}, {York},
  {Gupta}, {Noterdaeme}  \& {Srianand}}{{Kulkarni} et~al.}{2022}]{Kulkarni2022}
{Kulkarni} V.~P.,  {Bowen} D.~V.,  {Straka} L.~A.,  {York} D.~G.,  {Gupta} N.,
  {Noterdaeme} P.,   {Srianand} R.,  2022, \mn@doi [\apj]
  {10.3847/1538-4357/ac5fab}, \href
  {https://ui.adsabs.harvard.edu/abs/2022ApJ...929..150K} {929, 150}

\bibitem[\protect\citeauthoryear{{Levesque}, {Kewley}  \& {Larson}}{{Levesque}
  et~al.}{2010}]{Levesque2010}
{Levesque} E.~M.,  {Kewley} L.~J.,   {Larson} K.~L.,  2010, \mn@doi [\aj]
  {10.1088/0004-6256/139/2/712}, \href
  {https://ui.adsabs.harvard.edu/abs/2010AJ....139..712L} {139, 712}

\bibitem[\protect\citeauthoryear{{Lopez} et~al.,}{{Lopez}
  et~al.}{2018}]{Lopez2018}
{Lopez} S.,  et~al., 2018, \mn@doi [\nat] {10.1038/nature25436}, \href
  {https://ui.adsabs.harvard.edu/abs/2018Natur.554..493L} {554, 493}

\bibitem[\protect\citeauthoryear{{Ma}, {Hopkins}, {Faucher-Gigu{\`e}re},
  {Zolman}, {Muratov}, {Kere{\v{s}}}  \& {Quataert}}{{Ma}
  et~al.}{2016}]{Ma2016}
{Ma} X.,  {Hopkins} P.~F.,  {Faucher-Gigu{\`e}re} C.-A.,  {Zolman} N.,
  {Muratov} A.~L.,  {Kere{\v{s}}} D.,   {Quataert} E.,  2016, \mn@doi [\mnras]
  {10.1093/mnras/stv2659}, \href
  {https://ui.adsabs.harvard.edu/abs/2016MNRAS.456.2140M} {456, 2140}

\bibitem[\protect\citeauthoryear{Martin, Shapley, Coil, Kornei, Bundy, Weiner,
  Noeske  \& Schiminovich}{Martin et~al.}{2012}]{Martin_2012}
Martin C.~L.,  Shapley A.~E.,  Coil A.~L.,  Kornei K.~A.,  Bundy K.,  Weiner
  B.~J.,  Noeske K.~G.,   Schiminovich D.,  2012, \mn@doi [The Astrophysical
  Journal] {10.1088/0004-637x/760/2/127}, \href
  {https://doi.org/10.1088%2F0004-637x%2F760%2F2%2F127} {760, 127}

\bibitem[\protect\citeauthoryear{{Mingozzi} et~al.,}{{Mingozzi}
  et~al.}{2020}]{Mingozzi2020}
{Mingozzi} M.,  et~al., 2020, \mn@doi [\aap] {10.1051/0004-6361/201937203},
  \href {https://ui.adsabs.harvard.edu/abs/2020A&A...636A..42M} {636, A42}

\bibitem[\protect\citeauthoryear{{Mortensen}, {Keerthi Vasan}, {Jones},
  {Faucher-Gigu{\`e}re}, {Sanders}, {Ellis}, {Leethochawalit}  \&
  {Stark}}{{Mortensen} et~al.}{2021}]{Mortensen2021}
{Mortensen} K.,  {Keerthi Vasan} G.~C.,  {Jones} T.,  {Faucher-Gigu{\`e}re}
  C.-A.,  {Sanders} R.~L.,  {Ellis} R.~S.,  {Leethochawalit} N.,   {Stark}
  D.~P.,  2021, \mn@doi [\apj] {10.3847/1538-4357/abfa11}, \href
  {https://ui.adsabs.harvard.edu/abs/2021ApJ...914...92M} {914, 92}

\bibitem[\protect\citeauthoryear{Ménard, Wild, Nestor, Quider, Zibetti, Rao
  \& Turnshek}{Ménard et~al.}{2011}]{Menard2011}
Ménard B.,  Wild V.,  Nestor D.,  Quider A.,  Zibetti S.,  Rao S.,   Turnshek
  D.,  2011, \mn@doi [Monthly Notices of the Royal Astronomical Society]
  {10.1111/j.1365-2966.2011.18227.x}, \href
  {https://doi.org/10.1111/j.1365-2966.2011.18227.x} {417, 801}

\bibitem[\protect\citeauthoryear{Nestor, Johnson, Wild, Ménard, Turnshek, Rao
  \& Pettini}{Nestor et~al.}{2011}]{Nestor_2011}
Nestor D.~B.,  Johnson B.~D.,  Wild V.,  Ménard B.,  Turnshek D.~A.,  Rao S.,
   Pettini M.,  2011, \mn@doi [Monthly Notices of the Royal Astronomical
  Society] {10.1111/j.1365-2966.2010.17865.x}, \href
  {http://dx.doi.org/10.1111/j.1365-2966.2010.17865.x} {412, 1559}

\bibitem[\protect\citeauthoryear{{Nielsen}, {Churchill}, {Kacprzak}  \&
  {Murphy}}{{Nielsen} et~al.}{2013a}]{Nielsen2013}
{Nielsen} N.~M.,  {Churchill} C.~W.,  {Kacprzak} G.~G.,   {Murphy} M.~T.,
  2013a, \mn@doi [\apj] {10.1088/0004-637X/776/2/114}, \href
  {http://adsabs.harvard.edu/abs/2013ApJ...776..114N} {776, 114}

\bibitem[\protect\citeauthoryear{Nielsen, Churchill  \& Kacprzak}{Nielsen
  et~al.}{2013b}]{Nielsen_2013}
Nielsen N.~M.,  Churchill C.~W.,   Kacprzak G.~G.,  2013b, \mn@doi [The
  Astrophysical Journal] {10.1088/0004-637x/776/2/115}, \href
  {https://doi.org/10.1088%2F0004-637x%2F776%2F2%2F115} {776, 115}

\bibitem[\protect\citeauthoryear{{Noterdaeme}, {Srianand}  \&
  {Mohan}}{{Noterdaeme} et~al.}{2010a}]{Noterdaeme_2010}
{Noterdaeme} P.,  {Srianand} R.,   {Mohan} V.,  2010a, \mn@doi [Monthly Notices
  of the Royal Astronomical Society] {10.1111/j.1365-2966.2009.16169.x}, \href
  {https://ui.adsabs.harvard.edu/abs/2010MNRAS.403..906N} {403, 906}

\bibitem[\protect\citeauthoryear{{Noterdaeme}, {Petitjean}, {Ledoux},
  {L{\'o}pez}, {Srianand}  \& {Vergani}}{{Noterdaeme}
  et~al.}{2010b}]{noterdaeme2010a}
{Noterdaeme} P.,  {Petitjean} P.,  {Ledoux} C.,  {L{\'o}pez} S.,  {Srianand}
  R.,   {Vergani} S.~D.,  2010b, \mn@doi [\aap] {10.1051/0004-6361/201015147},
  \href {http://adsabs.harvard.edu/abs/2010A%26A...523A..80N} {523, A80}

\bibitem[\protect\citeauthoryear{{Putman}, {Peek}  \& {Joung}}{{Putman}
  et~al.}{2012}]{Putman2012}
{Putman} M.~E.,  {Peek} J.~E.~G.,   {Joung} M.~R.,  2012, \mn@doi [\araa]
  {10.1146/annurev-astro-081811-125612}, \href
  {https://ui.adsabs.harvard.edu/abs/2012ARA&A..50..491P} {50, 491}

\bibitem[\protect\citeauthoryear{{Rao}, {Belfort-Mihalyi}, {Turnshek},
  {Monier}, {Nestor}  \& {Quider}}{{Rao} et~al.}{2011}]{rao2011}
{Rao} S.~M.,  {Belfort-Mihalyi} M.,  {Turnshek} D.~A.,  {Monier} E.~M.,
  {Nestor} D.~B.,   {Quider} A.,  2011, \mn@doi [\mnras]
  {10.1111/j.1365-2966.2011.19119.x}, \href
  {http://adsabs.harvard.edu/abs/2011MNRAS.416.1215R} {416, 1215}

\bibitem[\protect\citeauthoryear{{Rigby}, {Charlton}  \& {Churchill}}{{Rigby}
  et~al.}{2002}]{Rigby2002}
{Rigby} J.~R.,  {Charlton} J.~C.,   {Churchill} C.~W.,  2002, \mn@doi [\apj]
  {10.1086/324723}, \href
  {https://ui.adsabs.harvard.edu/abs/2002ApJ...565..743R} {565, 743}

\bibitem[\protect\citeauthoryear{{Rubin}, {Prochaska}, {Koo}, {Phillips},
  {Martin}  \& {Winstrom}}{{Rubin} et~al.}{2014}]{Rubin2014}
{Rubin} K. H.~R.,  {Prochaska} J.~X.,  {Koo} D.~C.,  {Phillips} A.~C.,
  {Martin} C.~L.,   {Winstrom} L.~O.,  2014, \mn@doi [The Astrophysical
  Journal] {10.1088/0004-637X/794/2/156}, \href
  {https://ui.adsabs.harvard.edu/abs/2014ApJ...794..156R} {794, 156}

\bibitem[\protect\citeauthoryear{{Rubin} et~al.,}{{Rubin}
  et~al.}{2022}]{Rubin2022}
{Rubin} K. H.~R.,  et~al., 2022, \mn@doi [\apj] {10.3847/1538-4357/ac7b88},
  \href {https://ui.adsabs.harvard.edu/abs/2022ApJ...936..171R} {936, 171}

\bibitem[\protect\citeauthoryear{{Samui}, {Subramanian}  \& {Srianand}}{{Samui}
  et~al.}{2008}]{Samui2008}
{Samui} S.,  {Subramanian} K.,   {Srianand} R.,  2008, \mn@doi [\mnras]
  {10.1111/j.1365-2966.2008.12932.x}, \href
  {https://ui.adsabs.harvard.edu/abs/2008MNRAS.385..783S} {385, 783}

\bibitem[\protect\citeauthoryear{{Sardane}, {Turnshek}  \& {Rao}}{{Sardane}
  et~al.}{2015}]{sardane2015}
{Sardane} G.~M.,  {Turnshek} D.~A.,   {Rao} S.~M.,  2015, \mn@doi [\mnras]
  {10.1093/mnras/stv1506}, \href
  {http://adsabs.harvard.edu/abs/2015MNRAS.452.3192S} {452, 3192}

\bibitem[\protect\citeauthoryear{Shapiro \& Field}{Shapiro \&
  Field}{1976}]{shapiro1976consequences}
Shapiro P.~R.,  Field G.~B.,  1976, The Astrophysical Journal, 205, 762

\bibitem[\protect\citeauthoryear{{Speagle}, {Steinhardt}, {Capak}  \&
  {Silverman}}{{Speagle} et~al.}{2014}]{Speagle2014}
{Speagle} J.~S.,  {Steinhardt} C.~L.,  {Capak} P.~L.,   {Silverman} J.~D.,
  2014, \mn@doi [\apjs] {10.1088/0067-0049/214/2/15}, \href
  {https://ui.adsabs.harvard.edu/abs/2014ApJS..214...15S} {214, 15}

\bibitem[\protect\citeauthoryear{{Srianand}}{{Srianand}}{1996}]{Srianand1996}
{Srianand} R.,  1996, \mn@doi [\apj] {10.1086/177179}, \href
  {https://ui.adsabs.harvard.edu/abs/1996ApJ...462..643S} {462, 643}

\bibitem[\protect\citeauthoryear{{Srianand}, {Gupta}, {Petitjean}, {Noterdaeme}
   \& {Saikia}}{{Srianand} et~al.}{2008}]{srianand2008}
{Srianand} R.,  {Gupta} N.,  {Petitjean} P.,  {Noterdaeme} P.,   {Saikia}
  D.~J.,  2008, \mn@doi [\mnras] {10.1111/j.1745-3933.2008.00558.x}, \href
  {http://adsabs.harvard.edu/abs/2008MNRAS.391L..69S} {391, L69}

\bibitem[\protect\citeauthoryear{{Straka}, {Whichard}, {Kulkarni}, {Bishof},
  {Bowen}, {Khare}  \& {York}}{{Straka} et~al.}{2013}]{straka2013}
{Straka} L.~A.,  {Whichard} Z.~L.,  {Kulkarni} V.~P.,  {Bishof} M.,  {Bowen}
  D.,  {Khare} P.,   {York} D.~G.,  2013, \mn@doi [\mnras]
  {10.1093/mnras/stt1798}, \href
  {http://adsabs.harvard.edu/abs/2013MNRAS.436.3200S} {436, 3200}

\bibitem[\protect\citeauthoryear{Straka et~al.,}{Straka
  et~al.}{2015}]{Straka_2015}
Straka L.~A.,  et~al., 2015, \mn@doi [Monthly Notices of the Royal Astronomical
  Society] {10.1093/mnras/stu2739}, 447, 3856

\bibitem[\protect\citeauthoryear{{Tremonti}, {Moustakas}  \&
  {Diamond-Stanic}}{{Tremonti} et~al.}{2007}]{Tremonti2007}
{Tremonti} C.~A.,  {Moustakas} J.,   {Diamond-Stanic} A. a.~M.,  2007, \mn@doi
  [The Astrophysical Journall] {10.1086/520083}, \href
  {https://ui.adsabs.harvard.edu/abs/2007ApJ...663L..77T} {663, L77}

\bibitem[\protect\citeauthoryear{{Tumlinson} et~al.,}{{Tumlinson}
  et~al.}{2011}]{Tumlinson2011}
{Tumlinson} J.,  et~al., 2011, \mn@doi [Science] {10.1126/science.1209840},
  \href {https://ui.adsabs.harvard.edu/abs/2011Sci...334..948T} {334, 948}

\bibitem[\protect\citeauthoryear{{Tumlinson} et~al.,}{{Tumlinson}
  et~al.}{2013}]{tumlinson2013}
{Tumlinson} J.,  et~al., 2013, \mn@doi [\apj] {10.1088/0004-637X/777/1/59},
  \href {http://adsabs.harvard.edu/abs/2013ApJ...777...59T} {777, 59}

\bibitem[\protect\citeauthoryear{Virtanen et~al.,}{Virtanen
  et~al.}{2020}]{scipy2020}
Virtanen P.,  et~al., 2020, \mn@doi [Nature Methods]
  {10.1038/s41592-019-0686-2}, \href {https://rdcu.be/b08Wh} {17, 261}

\bibitem[\protect\citeauthoryear{{Wild}, {Hewett}  \& {Pettini}}{{Wild}
  et~al.}{2007}]{Wild2007}
{Wild} V.,  {Hewett} P.~C.,   {Pettini} M.,  2007, \mn@doi [\mnras]
  {10.1111/j.1365-2966.2006.11146.x}, \href
  {https://ui.adsabs.harvard.edu/abs/2007MNRAS.374..292W} {374, 292}

\bibitem[\protect\citeauthoryear{{York} et~al.,}{{York}
  et~al.}{2006}]{york2006}
{York} D.~G.,  et~al., 2006, \mn@doi [\mnras]
  {10.1111/j.1365-2966.2005.10018.x}, \href
  {http://adsabs.harvard.edu/abs/2006MNRAS.367..945Y} {367, 945}

\bibitem[\protect\citeauthoryear{{York} et~al.,}{{York}
  et~al.}{2012}]{york2012}
{York} D.~G.,  et~al., 2012, \mn@doi [\mnras]
  {10.1111/j.1365-2966.2012.21166.x}, \href
  {http://adsabs.harvard.edu/abs/2012MNRAS.423.3692Y} {423, 3692}

\bibitem[\protect\citeauthoryear{{Zhou} et~al.,}{{Zhou}
  et~al.}{2021}]{Zhou2021}
{Zhou} R.,  et~al., 2021, \mn@doi [\mnras] {10.1093/mnras/staa3764}, \href
  {https://ui.adsabs.harvard.edu/abs/2021MNRAS.501.3309Z} {501, 3309}

\bibitem[\protect\citeauthoryear{{Zhu} \& {M{\'e}nard}}{{Zhu} \&
  {M{\'e}nard}}{2013a}]{Zhu2013}
{Zhu} G.,  {M{\'e}nard} B.,  2013a, \mn@doi [\apj]
  {10.1088/0004-637X/770/2/130}, \href
  {https://ui.adsabs.harvard.edu/abs/2013ApJ...770..130Z} {770, 130}

\bibitem[\protect\citeauthoryear{{Zhu} \& {M{\'e}nard}}{{Zhu} \&
  {M{\'e}nard}}{2013b}]{Zhu2013ca2}
{Zhu} G.,  {M{\'e}nard} B.,  2013b, \mn@doi [\apj]
  {10.1088/0004-637X/773/1/16}, \href
  {https://ui.adsabs.harvard.edu/abs/2013ApJ...773...16Z} {773, 16}

\makeatother
\end{thebibliography}


\appendix
\label{sec:appendix}

\section{GOTOQ with photometric extensions}
\begin{table*}
    \centering
    \caption{GOTOQ with photometric extensions and their properties.{ Fluxes and the associated errors are measured in nanomaggies.}}
    \begin{tabular}{lcccccccccccr}
    \hline
    GOTOQ & RA & Dec & Plate & Fiber & MJD & $z_{qso}$ & $z_{abs}$ & D (kpc) & flux(nm) & error(nm) & $z_{ph}$ & $z_{ph}^{err}$\\
    \hline
    J0009+1107  &  2.291708    &  11.121000  &  6113  &  0602  &  56219  &  2.6470  &  0.6804  &  11.2  &  1.32  &  0.05  &  0.557  &  0.130\\
J0044+1524  &  11.114625   &  15.410889  &  6198  &  0086  &  56211  &  2.6090  &  0.7142  &  7.0   &  0.45  &  0.05  &  0.834  &  0.500\\
J0119+0505  &  19.974708   &  5.097472   &  4425  &  0122  &  55864  &  2.2000  &  0.4485  &  8.8   &  0.98  &  0.05  &  0.444  &  0.102\\
J0122+2736  &  20.699042   &  27.615833  &  6261  &  0732  &  56219  &  2.2950  &  0.7849  &  12.4  &  0.54  &  0.04  &  0.722  &  0.243\\
J0129-0032  &  22.317250   &  -0.536500  &  4229  &  0182  &  55501  &  1.8370  &  0.8991  &  13.5  &  2.46  &  0.04  &  1.069  &  0.173\\
J0131-0401  &  22.887167   &  -4.027083  &  7048  &  0710  &  56575  &  2.2700  &  0.5017  &  10.7  &  2.21  &  0.03  &  0.395  &  0.095\\
J0235+0059  &  38.874250   &  0.987861   &  3744  &  0952  &  55209  &  2.7160  &  0.9250  &  11.6  &  0.28  &  0.02  &  0.943  &  0.204\\
J0236-0005  &  39.079083   &  -0.091417  &  0408  &  0237  &  51821  &  0.9806  &  0.6448  &  8.6   &  0.73  &  0.02  &  0.970  &  0.204\\
J0748+1656  &  117.210792  &  16.940417  &  1920  &  0217  &  53314  &  0.9321  &  0.6651  &  15.0  &  1.09  &  0.07  &  0.961  &  0.147\\
J0753+3603$^\star$  &  118.395458  &  36.061250  &  3791  &  0894  &  55501  &  0.5260  &  0.3994  &  14.5  &  1.23  &  0.05  &  0.424  &  0.160\\
J0807+0945  &  121.962625  &  9.765194   &  4510  &  0081  &  55559  &  2.5900  &  0.7470  &  16.3  &  1.33  &  0.05  &  0.568  &  0.074\\
J0808+0641  &  122.036125  &  6.685750   &  1756  &  0184  &  53080  &  2.1133  &  0.4326  &  9.2   &  2.08  &  0.07  &  0.385  &  0.120\\
J0810+3328$^\star$  &  122.562417  &  33.475056  &  3757  &  0175  &  55508  &  2.2160  &  0.4643  &  17.8  &  2.84  &  0.08  &  0.470  &  0.076\\
J0826+5533  &  126.580000  &  55.556528  &  5151  &  0158  &  56567  &  2.3930  &  0.7986  &  14.7  &  0.86  &  0.10  &  0.691  &  0.165\\
J0839+3853  &  129.977792  &  38.898333  &  3765  &  0606  &  55508  &  1.9630  &  0.6085  &  6.5   &  0.42  &  0.08  &  0.960  &  0.234\\
J0847+3420  &  131.969208  &  34.337667  &  5186  &  0632  &  56337  &  2.4970  &  0.7371  &  13.6  &  0.81  &  0.11  &  0.554  &  0.157\\
J0855+1933  &  133.950750  &  19.558556  &  5175  &  0690  &  55955  &  1.9270  &  0.8849  &  8.3   &  1.01  &  0.13  &  0.913  &  0.574\\
J0858+0556  &  134.627208  &  5.935000   &  1191  &  0283  &  52674  &  1.8524  &  0.4318  &  7.0   &  1.83  &  0.05  &  0.378  &  0.117\\
J0920+3133  &  140.142917  &  31.565361  &  2961  &  0082  &  54550  &  1.7232  &  0.8035  &  9.8   &  0.36  &  0.04  &  0.963  &  0.264\\
J0929-0126  &  142.312583  &  -1.439167  &  3781  &  0978  &  55243  &  2.2500  &  0.6990  &  13.4  &  0.59  &  0.06  &  0.968  &  0.214\\
J0930+0018  &  142.585833  &  0.307778   &  3823  &  0996  &  55534  &  2.4300  &  0.5928  &  5.9   &  0.77  &  0.06  &  0.473  &  0.213\\
J0930+1139$^\star$  &  142.730833  &  11.665722  &  5313  &  0376  &  55973  &  0.7860  &  0.5524  &  19.5  &  4.39  &  0.07  &  0.543  &  0.062\\
J0935+5205$^\star$  &  143.865083  &  52.096722  &  0768  &  0253  &  52281  &  2.5787  &  0.6717  &  17.8  &  0.69  &  0.11  &  0.726  &  0.284\\
J0951+1116  &  147.992375  &  11.279361  &  5321  &  0844  &  55945  &  2.3420  &  0.7564  &  12.7  &  2.94  &  0.11  &  0.656  &  0.062\\
J0954+2017  &  148.536250  &  20.294639  &  5785  &  0052  &  56269  &  2.2300  &  0.5488  &  5.9   &  1.40  &  0.07  &  0.602  &  0.150\\
J0959+4809  &  149.828042  &  48.161944  &  1006  &  0248  &  52708  &  1.2848  &  0.7745  &  13.2  &  1.12  &  0.10  &  1.032  &  0.152\\
J1000+4438  &  150.064583  &  44.646639  &  0942  &  0511  &  52703  &  1.8778  &  0.7185  &  9.8   &  1.28  &  0.08  &  0.815  &  0.277\\
J1021+2152  &  155.413333  &  21.878917  &  5872  &  0916  &  56027  &  2.2020  &  0.7601  &  9.5   &  0.37  &  0.07  &  0.999  &  0.307\\
J1023+1427  &  155.953208  &  14.462444  &  5340  &  0414  &  56011  &  2.3080  &  0.7142  &  11.7  &  0.37  &  0.06  &  0.675  &  0.249\\
J1031+3158  &  157.786958  &  31.974639  &  6451  &  0640  &  56358  &  2.4930  &  0.5738  &  10.8  &  0.63  &  0.07  &  0.473  &  0.097\\
J1040+0151  &  160.185667  &  1.856139   &  4734  &  0108  &  55646  &  2.1420  &  0.4455  &  11.2  &  1.77  &  0.07  &  0.518  &  0.082\\
J1041+3101  &  160.477000  &  31.029917  &  2019  &  0060  &  53430  &  0.8559  &  0.7858  &  10.5  &  1.35  &  0.08  &  0.817  &  0.141\\
J1042+3226$^\star$  &  160.656417  &  32.437750  &  6447  &  0084  &  56362  &  2.3500  &  0.6946  &  18.4  &  5.18  &  0.09  &  0.726  &  0.057\\
J1056+2432  &  164.044125  &  24.539833  &  6418  &  0604  &  56354  &  1.1400  &  0.3981  &  16.9  &  5.08  &  0.08  &  0.414  &  0.135\\
J1101+1029  &  165.106000  &  10.488500  &  5361  &  0510  &  55973  &  2.5460  &  0.8563  &  12.1  &  0.47  &  0.05  &  0.744  &  0.357\\
J1102+3159  &  165.576000  &  31.993111  &  6442  &  0134  &  56369  &  2.5630  &  0.7168  &  7.0   &  0.18  &  0.06  &  0.519  &  0.473\\
J1117+0759  &  169.482125  &  7.992278   &  5369  &  0393  &  56272  &  2.6350  &  0.7681  &  8.4   &  0.32  &  0.04  &  0.568  &  0.151\\
J1119+5901  &  169.849208  &  59.022278  &  7108  &  0291  &  56686  &  1.3930  &  0.5316  &  12.1  &  2.59  &  0.09  &  0.354  &  0.099\\
J1130+1245  &  172.566250  &  12.765861  &  1606  &  0518  &  53055  &  1.5024  &  0.5330  &  10.5  &  1.53  &  0.07  &  0.430  &  0.112\\
J1131+2021  &  172.784792  &  20.364194  &  2502  &  0371  &  54180  &  1.7682  &  0.5631  &  7.0   &  4.10  &  0.08  &  0.404  &  0.234\\
J1159+3935  &  179.861458  &  39.585167  &  2027  &  0570  &  53433  &  1.6658  &  0.4921  &  11.4  &  3.02  &  0.09  &  0.450  &  0.176\\
J1215+6135  &  183.908333  &  61.590306  &  6972  &  0896  &  56426  &  2.3150  &  0.6622  &  7.4   &  0.60  &  0.07  &  0.823  &  0.200\\
J1218+0309  &  184.658750  &  3.150028   &  4750  &  0774  &  55630  &  2.2510  &  0.5222  &  11.4  &  2.57  &  0.07  &  0.445  &  0.063\\
J1219+6707  &  184.926667  &  67.128611  &  6975  &  0904  &  56720  &  2.4790  &  0.8029  &  11.8  &  0.51  &  0.08  &  0.953  &  0.246\\
J1230+3652  &  187.528458  &  36.874556  &  3965  &  0070  &  55302  &  2.1200  &  0.8109  &  9.7   &  0.80  &  0.08  &  0.952  &  0.197\\
J1235+0304  &  188.809833  &  3.071306   &  4754  &  0592  &  55649  &  2.1950  &  0.5313  &  13.0  &  4.04  &  0.06  &  0.485  &  0.054\\
J1243+5203$^\star$  &  190.881917  &  52.059750  &  6674  &  0926  &  56416  &  2.1730  &  0.7748  &  20.0  &  0.33  &  0.06  &  0.916  &  0.246\\
J1253+4532  &  193.266292  &  45.538694  &  7417  &  0274  &  56753  &  1.3320  &  0.4424  &  5.9   &  1.72  &  0.08  &  0.781  &  0.229\\
J1253+1758  &  193.412917  &  17.975556  &  2601  &  0625  &  54144  &  0.5042  &  0.4008  &  8.5   &  9.28  &  0.28  &  0.494  &  0.132\\
J1312+4006  &  198.181667  &  40.108583  &  4706  &  0156  &  55705  &  2.4060  &  0.7293  &  8.9   &  0.23  &  0.09  &  0.844  &  0.165\\
J1314+0330  &  198.725833  &  3.511750   &  4760  &  0096  &  55656  &  2.7000  &  0.7792  &  13.5  &  0.43  &  0.07  &  0.953  &  0.360\\
J1346+1236  &  206.606542  &  12.604028  &  1701  &  0401  &  53142  &  1.5233  &  0.7120  &  8.0   &  1.81  &  0.09  &  0.704  &  0.179\\
J1354+3217$^\star$  &  208.623625  &  32.285806  &  3861  &  0314  &  55274  &  0.6830  &  0.6398  &  21.8  &  3.81  &  0.07  &  0.523  &  0.180\\
J1354+0756  &  208.690500  &  7.943778   &  1806  &  0043  &  53559  &  1.3235  &  0.7126  &  14.2  &  1.74  &  0.07  &  0.838  &  0.116\\
J1356+0601  &  209.142333  &  6.026000   &  1805  &  0013  &  53875  &  1.3929  &  0.6292  &  11.2  &  1.96  &  0.05  &  0.613  &  0.091\\
J1416+2214$^\star$  &  214.017167  &  22.235917  &  2786  &  0440  &  54540  &  1.1180  &  0.6995  &  21.1  &  4.63  &  0.08  &  0.650  &  0.079\\
J1417+4730  &  214.490583  &  47.502500  &  6751  &  0824  &  56368  &  1.2410  &  0.9111  &  13.0  &  0.58  &  0.10  &  0.920  &  0.252\\
J1422+3222  &  215.608875  &  32.372528  &  1840  &  0202  &  53472  &  1.6295  &  0.5413  &  11.3  &  2.44  &  0.08  &  0.486  &  0.094\\
J1430+5420  &  217.565125  &  54.338361  &  6710  &  0262  &  56416  &  2.7870  &  1.0582  &  10.0  &  0.21  &  0.07  &  0.988  &  0.239\\
J1435+4740  &  218.793000  &  47.682944  &  1673  &  0023  &  53462  &  0.7844  &  0.4407  &  11.3  &  2.12  &  0.09  &  0.443  &  0.192\\
J1438+3522  &  219.501958  &  35.369000  &  3865  &  0702  &  55272  &  2.0530  &  0.7869  &  11.7  &  0.39  &  0.09  &  0.952  &  0.206\\
J1442+6009$^\star$  &  220.630208  &  60.165083  &  0607  &  0037  &  52368  &  2.0399  &  0.7987  &  21.8  &  3.48  &  0.13  &  0.757  &  0.126\\
\hline
\end{tabular}
\end{table*}

\begin{table*}
    \ContinuedFloat
    \centering
    \caption{Continued}
    \begin{tabular}{lcccccccccccr}
    \hline
    GOTOQ & RA & Dec & Plate & Fiber & MJD & $z_{qso}$ & $z_{abs}$ & D (kpc) & flux(nm) & error(nm) & $z_{ph}$ & $z_{ph}^{err}$\\
    \hline
    J1449+2136  &  222.490083  &  21.609139  &  5905  &  0696  &  56065  &  2.6640  &  0.7977  &  8.5   &  0.76  &  0.07  &  0.634  &  0.717\\
J1454+4424  &  223.717667  &  44.402528  &  6046  &  0933  &  56096  &  2.3960  &  0.6630  &  10.8  &  1.00  &  0.11  &  0.621  &  0.249\\
J1502+1806  &  225.567375  &  18.115944  &  3957  &  0494  &  55664  &  1.1110  &  0.5442  &  10.9  &  2.80  &  0.08  &  0.474  &  0.109\\
J1509+2630  &  227.368750  &  26.500778  &  3872  &  0246  &  55382  &  2.9370  &  0.7790  &  12.7  &  0.62  &  0.05  &  0.821  &  0.143\\
J1509+1252  &  227.400375  &  12.879139  &  2752  &  0234  &  54533  &  1.9031  &  0.5405  &  6.8   &  1.55  &  0.06  &  0.444  &  0.151\\
J1518+5815  &  229.741583  &  58.257639  &  0612  &  0562  &  52079  &  1.7390  &  0.6470  &  12.2  &  3.65  &  0.11  &  0.458  &  0.203\\
J1525+2647  &  231.260958  &  26.791861  &  3959  &  0466  &  55679  &  2.2420  &  0.9126  &  13.9  &  0.63  &  0.09  &  0.698  &  0.343\\
J1530+4152  &  232.515208  &  41.874528  &  6049  &  0166  &  56091  &  2.3910  &  0.8468  &  12.2  &  0.86  &  0.06  &  1.127  &  0.228\\
J1634+4849  &  248.699500  &  48.829139  &  6318  &  0555  &  56186  &  2.4960  &  0.7477  &  14.5  &  0.32  &  0.06  &  0.730  &  0.200\\
J1643+1729  &  250.868500  &  17.498722  &  4062  &  0130  &  55383  &  2.1920  &  0.5442  &  9.5   &  0.89  &  0.04  &  0.590  &  0.299\\
J1646+2733$^\star$  &  251.678708  &  27.560889  &  1690  &  0439  &  53475  &  1.6289  &  0.7985  &  17.5  &  0.71  &  0.10  &  0.833  &  0.258\\
J1656+4146  &  254.133917  &  41.771472  &  6037  &  0980  &  56074  &  2.1630  &  0.6622  &  12.2  &  2.72  &  0.13  &  0.906  &  0.348\\
J1657+2045  &  254.353417  &  20.766528  &  1425  &  0358  &  52913  &  1.4585  &  0.6254  &  10.8  &  1.38  &  0.14  &  0.416  &  0.434\\
J1658+2940  &  254.742708  &  29.682083  &  5013  &  0664  &  55723  &  2.5770  &  0.8096  &  7.8   &  0.27  &  0.05  &  1.026  &  0.543\\
J2147+1048  &  326.874958  &  10.808556  &  0733  &  0288  &  52207  &  2.1097  &  0.6990  &  12.0  &  0.78  &  0.09  &  0.668  &  0.146\\
J2150+0225  &  327.746208  &  2.432556   &  5146  &  0878  &  55831  &  2.3780  &  0.5682  &  6.9   &  0.45  &  0.04  &  0.919  &  0.194\\
J2219+0239  &  334.960417  &  2.664750   &  4319  &  0286  &  55507  &  2.5280  &  0.7203  &  10.4  &  0.56  &  0.04  &  0.548  &  0.162\\
J2244-0215  &  341.021208  &  -2.263306  &  4363  &  0778  &  55537  &  2.5740  &  0.9243  &  16.3  &  1.22  &  0.05  &  1.059  &  0.097\\
J2253+2117  &  343.360667  &  21.290861  &  6120  &  0966  &  56206  &  2.4580  &  0.6994  &  6.8   &  0.76  &  0.08  &  0.864  &  0.960\\
J2324+2427  &  351.025042  &  24.464472  &  6304  &  0456  &  56570  &  2.2780  &  0.8923  &  8.7   &  0.12  &  0.05  &  0.961  &  0.338\\
J2339+1540  &  354.916792  &  15.668278  &  6138  &  0488  &  56598  &  2.3680  &  0.7042  &  11.7  &  0.48  &  0.05  &  0.575  &  0.185\\
J2342+0101  &  355.739583  &  1.022556   &  4213  &  0912  &  55449  &  2.4530  &  0.6768  &  9.5   &  0.66  &  0.02  &  0.437  &  0.358\\
\hline

\end{tabular}
\label{tab:extended_GOTOQ}
\end{table*}

\bsp	
\label{lastpage}
\end{document}